\PassOptionsToPackage{table}{xcolor}
\documentclass[authorversion]{article}
\usepackage{iclr2025_conference,times}

\usepackage{amsmath,amsfonts,bm}

\def\eqref#1{equation~\ref{#1}}

\def\1{\bm{1}}

\DeclareMathAlphabet{\mathsfit}{\encodingdefault}{\sfdefault}{m}{sl}
\SetMathAlphabet{\mathsfit}{bold}{\encodingdefault}{\sfdefault}{bx}{n}

\usepackage{array}
\newcolumntype{H}{>{\setbox0=\hbox\bgroup}c<{\egroup}@{}}

\usepackage{adjustbox}
\usepackage{color}
\usepackage{xcolor}
\usepackage{longtable}
\usepackage{listings}
\usepackage{lineno}
\usepackage{hyperref}
\usepackage{url}
\usepackage{enumitem}
\usepackage{xspace}
\usepackage{enumitem}
\usepackage{bbm}
\usepackage{balance}
\usepackage{caption}
\usepackage{subcaption}
\usepackage{tikz, pgfplots}
\usepackage{wrapfig}
\usepackage{verbatimbox}
\usepackage{pgfplotstable}
\usepackage{multirow}
\usepackage{enumitem}
\usepackage{stackengine}
\usepackage{scalerel}
\usepackage{booktabs}
\usepackage{diagbox}
\usepackage{pifont}
\usepackage{algorithmicx}
\usepackage{algorithm}
\usepackage{amssymb}
\usepackage{amsmath}

\usepackage[multiple]{footmisc}
\usetikzlibrary{patterns}
\usepackage{tcolorbox}
\tcbuselibrary{listings,skins}
\usepackage{ifthen}

\definecolor{WowColor}{rgb}{.75,0,.75}
\definecolor{SubtleColor}{rgb}{0,0,.50}
\newcommand{\cmark}{\ding{51}\xspace}%
\newcommand{\xmark}{\ding{55}\xspace}%

\ifdefined\Comment
        \renewcommand{\Comment}[1]{}
\else
        \newcommand{\Comment}[1]{}
\fi
\newcommand{\NA}[1]{\textcolor{SubtleColor}{ {\tiny \bf ($\star$)} #1}}
\newcommand{\Fix}[1]{\textcolor{red}{[#1]}}

\newcounter{margincounter}

\newcommand{\mypara}[1]{\vspace{.03in}\noindent \textbf{#1.}}

\definecolor{ghgreen}{rgb}{0.90,1,0.93}
\definecolor{ghred}{rgb}{1,0.88,0.94}

\definecolor{codegreen}{rgb}{0,0.6,0}
\definecolor{codegray}{rgb}{0.5,0.5,0.5}
\definecolor{codepurple}{rgb}{0.58,0,0.82}
\definecolor{backcolour}{rgb}{0.95,0.95,0.92}

\definecolor{codegreen}{rgb}{0,0.6,0}
\definecolor{codegray}{rgb}{0.5,0.5,0.5}
\definecolor{codepurple}{rgb}{0.58,0,0.82}
\definecolor{backcolour}{rgb}{0.95,0.95,0.92}

\lstdefinestyle{codeqlstyle}{
    commentstyle=\color{codegreen},
    keywordstyle=\color{magenta},
    numberstyle=\tiny\color{codegray},
    stringstyle=\color{codepurple},
    basicstyle=\footnotesize\ttfamily,
    breakatwhitespace=false,         
    breaklines=true,                 
    captionpos=b,                    
    keepspaces=true,                 
    numbers=left,                    
    numbersep=5pt,                  
    showspaces=false,                
    showstringspaces=false,
    showtabs=false,                  
    tabsize=2,
    language=SQL,
    morecomment=[l]{//},%
    morekeywords={select, from, where, and, or, not, predicate, class, extends, import, module, with, without, string,bindingset,if}, 
}

\lstdefinelanguage{MyPrompt}{
  keywords={System, Response, User},
  sensitive = true,
  comment=[l]{//}, 
  morecomment=[s]{/*}{*/}
}
\lstdefinestyle{mypromptstyle}{
    language=MyPrompt,  
    backgroundcolor=\color{backcolour},
    basicstyle=\footnotesize\ttfamily,
    commentstyle=\color{codegreen},
    keywordstyle=\color{codepurple}\bfseries,
    numberstyle=\tiny\color{codegray},
    stringstyle=\color{blue},    
    breakatwhitespace=false,
    breaklines=true,
    captionpos=b,
    keepspaces=true,
    numbers=left,
    numbersep=5pt,
    tabsize=4,
    columns=fullflexible,
    morecomment=[s]{[}{]},
}

\newif\ifNIPSSUB
\newcommand{\ifnotNIPSSUB}{\ifNIPSSUB\else}
\newcommand{\tool}{IRIS\xspace}
\newcommand{\benchmark}{CWE-Bench-Java\xspace}

\newcommand{\vulc}{vulnerability class\xspace}

\definecolor{checkgreen}{rgb}{0.12, 0.632, 0.01}
\newcommand{\green}[1]{{\color{checkgreen} #1}}

\definecolor{checkred}{rgb}{0.624, 0.1, 0.01}
\newcommand{\red}[1]{{\color{checkred} #1}}

\newcommand{\todo}[1]{{\color{red} (todo: #1)}}
\newcommand{\mayur}[1]{{\color{blue} (MN: #1)}}

\newcommand{\totalvuls}{120\xspace}

\iclrfinalcopy
\begin{document}

\title{
IRIS: LLM-Assisted Static Analysis for \\ Detecting Security Vulnerabilities}

\ifnotNIPSSUB
\author{%
  Ziyang Li \\
  University of Pennsylvania \\
  \texttt{liby99@cis.upenn.edu} \\
  \And
  Saikat Dutta \\
  Cornell University \\
  \texttt{saikatd@cornell.edu} \\
  \And
  Mayur Naik \\
  University of Pennsylvania \\
  \texttt{mhnaik@cis.upenn.edu}
}
\fi
\maketitle
\vspace{-0.2in}
\begin{abstract}
Software is prone to security vulnerabilities.
Program analysis tools to detect them have limited effectiveness in practice due to their reliance on human labeled specifications.
Large language models (or LLMs) have shown impressive code generation capabilities but they cannot do complex reasoning over code to detect such vulnerabilities especially since this task requires whole-repository analysis.
We propose \tool, a neuro-symbolic approach that systematically combines LLMs with static analysis to perform whole-repository reasoning for security vulnerability detection.
Specifically, \tool leverages LLMs to infer taint specifications and perform contextual analysis, alleviating needs for human specifications and inspection.
For evaluation, we curate a new dataset, \benchmark, comprising 120 manually validated security vulnerabilities in real-world Java projects.
A state-of-the-art static analysis tool CodeQL detects only 27 of these vulnerabilities whereas \tool with GPT-4 detects 55 ($+28$) and improves upon CodeQL's average false discovery rate by 5\% points.
Furthermore, IRIS identifies 4 previously unknown vulnerabilities which cannot be found by existing tools. \tool is available publicly at \url{https://github.com/iris-sast/iris}.



%

\end{abstract}




\section{Introduction}

Security vulnerabilities pose a major threat to the safety of software applications and its users.
In 2023 alone, more than 29,000 CVEs were reported---almost 4000 higher than in 2022~\citep{cvedetails}.
Detecting vulnerabilities is extremely challenging despite advances in techniques to uncover them.
A promising such technique called static taint analysis is widely used in popular tools such as GitHub CodeQL~\citep{codeql}, Facebook Infer~\citep{fbinfer},
Checker Framework~\citep{checker}, and Snyk Code~\citep{synkio}. These tools, however, face several challenges that greatly limit their effectiveness and accessibility in practice.

\begin{figure}[!bh]
\vspace{-0px}
\centering
\includegraphics[width=0.98\textwidth]{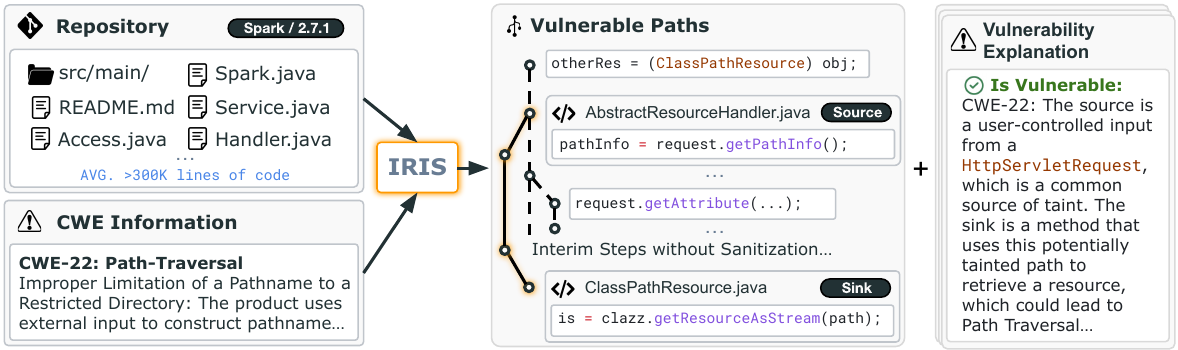}
\caption{
Overview of the \tool neuro-symbolic system.
It checks a given whole repository for a given type of vulnerability (CWE) and outputs a set of potential vulnerable paths with explanations.
}
\label{fig:pipeline}
\end{figure}


\mypara{False negatives due to missing taint specifications of third-party library APIs}
First, static
taint analysis predominantly relies on \emph{specifications} of third-party
library APIs as sources, sinks, or sanitizers. 
In practice, developers and
analysis engineers have to manually craft such specifications based on their
domain knowledge and API documentation. 
This is a laborious and
error-prone process that often leads to missing specifications and incomplete
analysis of vulnerabilities. Further, even if such specifications may exist for
many libraries, they need to be periodically updated to capture changes in newer
versions of such libraries and also cover new libraries that are developed.

\mypara{False positives due to lack of precise context-sensitive and intuitive reasoning}
Second, it is well-known that static analysis often suffers from low precision,
i.e., it may generate many false alarms~\citep{kang2022detecting,johnson2013don}. 
Such imprecision stems from
multiple sources. For instance, the source or sink specifications may be
spurious, or the analysis may over-approximate over branches in code or possible inputs. Further, even if the specifications are correct, the
context in which the detected source or sink is used may not be exploitable. Hence, a developer may need to triage through several 
potentially false security alerts, wasting significant time and effort.
\Comment{Moreover, to triage each detected alert, a developer needs a combination of
different skills: domain expertise to understand the nature of the project and
and third-party libraries used, security expertise to assess the security
implications of a detected alert, and program analysis expertise to effectively use the
static analysis tool and distinguish between spurious and malicious alerts -- all of
which makes triaging alerts an extremely challenging and labor-intensive endeavor.}

\mypara{Limitations of prior data-driven approaches to improve static taint analysis} 
Many techniques have been proposed to address the challenges
of static taint analysis. 
For instance, \cite{livshits2009merlin} proposed a probabilistic approach, MERLIN, to automatically mine taint specifications.
A more recent work, Seldon~\citep{chibotaru2019scalable}, improves the scalability of this approach by formulating the taint specification inference problem as a linear optimization task. 
However, such approaches rely on analyzing the code of third-party libraries to extract specifications, which is expensive and hard to scale. 
Researchers have also developed statistical and learning-based techniques to mitigate false positive
alerts~\citep{jung2005taming,heckman2009model,ranking2014finding}. However,
such approaches still have limited effectiveness in practice~\citep{kang2022detecting}. 

Large Language Models (or LLMs) have made impressive strides in code generation and summarization. LLMs have also been applied to code related tasks such as
program repair~\citep{xia2023automated}, code translation~\citep{pan2024lost},
test generation~\citep{lemieux2023codamosa}, and static
analysis~\citep{li2024enhancing}. Recent
studies~\citep{steenhoek2024comprehensive,khare2023understanding} evaluated LLMs'
effectiveness at detecting vulnerabilities at the method level and showed that
LLMs fail to do complex reasoning with code, especially because it depends on
the \emph{context} in which the method is used in the project. On the other
hand, recent benchmarks like SWE-Bench~\citep{jimenez2023swe} show that LLMs are
also poor at doing project-level reasoning.
Hence, an intriguing question is
whether LLMs can be combined with static analysis to improve their reasoning capabilities.
In this work, we answer this question in the context of vulnerability detection and make the following contributions:

\mypara{Approach} We propose \textbf{\tool}, a neuro-symbolic approach for vulnerability detection that combines the strengths of static analysis and LLMs.
Fig.~\ref{fig:pipeline} presents an overview of \tool.
Given a project to analyze for a given \vulc (or CWE), \tool applies LLMs for mining CWE-specific taint specifications. 
\tool augments such specifications with CodeQL, a tool for static taint analysis. 
Our intuition here is because LLMs have seen numerous usages of such library APIs, they have an understanding of the relevant APIs for different CWEs.
Further, to address the imprecision problem of static analysis, we propose a contextual analysis technique with LLMs that reduces the false positive alarms and minimizes the triaging effort for developers. 
Our key insight is that encoding the code-context and path-sensitive information in the prompt elicits more reliable reasoning from LLMs. 
Finally, our neuro-symbolic approach allows LLMs to do more precise whole-repository reasoning and minimizes the human effort involved in using static analysis tools.

\mypara{Dataset} We curate a dataset of manually vetted and compilable Java projects, \textbf{\benchmark}, 
containing \totalvuls vulnerabilities (one per project) across four common vulnerability classes. 
The projects in the dataset are complex, containing 300K lines of code on average, and 10 projects with more than a million lines of code each, making it a challenging benchmark for vulnerability detection.
The dataset and the corresponding scripts to fetch, build, and analyze the Java projects are available publicly at \url{https://github.com/iris-sast/cwe-bench-java}.


\mypara{Results} We evaluate \tool on \benchmark 
using 7 diverse open- and closed-source LLMs. Overall, \tool obtains the best results with GPT-4, detecting
55 vulnerabilities, which is 28 ($103.7\%$) more than CodeQL, the existing best-performing static analyzer.
We show that the increase is not at the expense of false positives, as IRIS with GPT-4 achieves an average false discovery rate of $84.82\%$, which is $5.21\%$ lower than that of CodeQL.
Further, when applied to the latest versions of 30 Java projects, IRIS with GPT-4 discovered 4 previously unknown vulnerabilities.

\section{Motivating Example}
\label{sec:motivating}

\begin{figure}[!t]
\centering
\includegraphics[width=\linewidth]{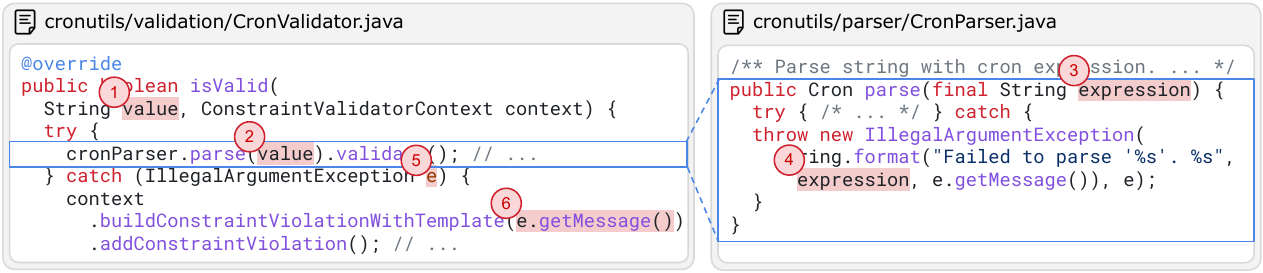}
\caption{
An example of Code Injection (CWE-94) vulnerability found in cron-utils (CVE-2021-41269) that CodeQL fails to detect. 
We number the program points of the vulnerable path.
}
\label{fig:motivating-vul}
\vspace{-10px}
\end{figure}

We illustrate the effectiveness of \tool in detecting a previously known code-injection (CWE-094) vulnerability in cron-utils (ver.~9.1.5),
a Java library for Cron data manipulation.
Fig.~\ref{fig:motivating-vul} shows the relevant code snippets.
A user-controlled string \texttt{value} passed into \texttt{isValid} function is transferred without sanitization to the \texttt{parse} function. 
If an exception is thrown, the function constructs an error message with the input.
However, the error message is used to invoke method \texttt{buildConstraintViolationWithTemplate} of class \texttt{ConstraintValidatorContext} in \texttt{javax.validator}, which interprets the message string as a Java Expression Language (Java EL) expression.
A malicious user may exploit this vulnerability by crafting a string containing a shell command 
such as \texttt{Runtime.exec(`rm -rf /')} to delete critical files on the server.

Detecting this vulnerability poses several challenges.
First, the cron-utils library consists of 13K SLOC (lines of code excluding blanks and comments), which needs to be analyzed to find this vulnerability.
%
This process requires analyzing data and control flow across several internal methods and third-party APIs.
Second, the analysis needs to identify relevant \emph{sources} and \emph{sinks}.
In this case, the \texttt{value} parameter of the public \texttt{isValid} method may contain arbitrary strings when invoked, and hence may be a source of malicious data. 
Additionally, external APIs like \texttt{buildConstraintViolationWithTemplate} can execute arbitrary Java EL expressions, hence they should be treated as sinks that are vulnerable to Code Injection attacks.
%
Finally, the analysis also requires identifying any sanitizers that block the flow of untrusted data.

Modern static analysis tools, like CodeQL, are effective at tracing taint data flows across complex codebases. However, CodeQL fails to detect this vulnerability due to missing specifications. CodeQL includes many manually curated specifications for sources and sinks across more than 360 popular Java library modules.
However, manually obtaining such specifications requires significant human effort to analyze, specify, and validate. 
Further, even with perfect specifications, CodeQL may often generate numerous false positives due to a lack of contextual reasoning, increasing the developer's burden of triaging the results.

In contrast, \tool takes a different approach by inferring project- and vulnerability-specific specifications \emph{on-the-fly} by using LLMs. 
The LLM-based components in \tool correctly identify the untrusted source and the vulnerable sink. 
\tool augments CodeQL with these specifications and successfully detects the unsanitized dataflow path between the detected source and sink in the repository.
However, augmented CodeQL produces many false positives, which are hard to eliminate using logical rules. 
To solve this challenge, \tool encodes the detected code paths and the surrounding context into a simple prompt and uses an LLM to classify it as true or false positive. 
Specifically, out of 8 paths reported by static analysis, 5 false positives are filtered out, leaving the path in Fig.~\ref{fig:motivating-vul} as one of the final alarms.
Overall, we observe that \tool can detect many such vulnerabilities that are beyond the reach of CodeQL-like static analysis tools, while keeping false alarms to a minimum.


\vspace{-0.05in}
\section{\tool Framework}


At a high level, \tool takes a Java project $P$, the
vulnerability class $C$ to detect, and a large language model $\texttt{LLM}$, as inputs.
\tool statically analyzes the project $P$, checks for vulnerabilities specific to $C$, and returns a set of potential security alerts $A$. 
Each alert is accompanied by a unique code path from a taint source to a taint sink that is vulnerable to $C$ (i.e., the path is unsanitized).

\ifNIPSSUB
\else
As illustrated in Fig.~\ref{fig:appendix-pipeline}, \tool has four main stages: 
First, \tool builds the given Java project and uses static analysis to extract all candidate APIs, including invoked external APIs and internal function parameters.
Second, \tool queries an LLM to label these APIs as sources or sinks that are specific to the given vulnerability class $C$. 
Third, \tool transforms the labeled sources and sinks into specifications that can be fed into a static analysis engine, such as CodeQL, and runs a \vulc-specific taint analysis query to detect vulnerabilities of that class in the project. 
This step generates a set of vulnerable code paths (or alerts) in the project. 
Finally, \tool triages the generated alerts by automatically filtering false positives, and presents them to the developer.

\begin{figure}[t]
    \includegraphics[width=\linewidth]{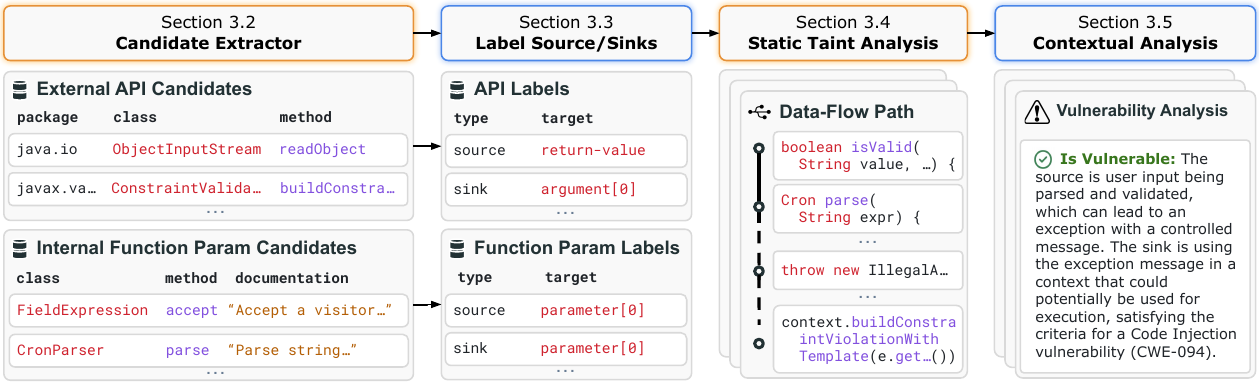}
    \caption{An illustration of the \tool\ pipeline.}
    \label{fig:appendix-pipeline}
    \vspace{-0.2in}
\end{figure}
\fi

\Comment{
Algorithm~\ref{algo:main} describes the main algorithm of \tool. \Fix{we can
remove the algo if we have workflow diagram.}\mayur{I propose to remove the algorithm and use a workflow diagram.} We next describe the main
components of \tool.

\begin{algorithm}
    \caption{\tool Algorithm}
    \label{algo:main}
    \begin{algorithmic}[1] 
        \State \textbf{Input:} Project $P$, Version $V$, Vulnerability Class $T$, LLM $M$
        \State \textbf{Output:} Alerts $A$
        \State $\textit{Sources}_C \gets \textit{GetCandidateSources}(P, V)$
        \State $\textit{Sinks}_C \gets \textit{GetCandidateSinks}(P, V)$
        \State $\textit{Sources}_L \gets \textit{LabelSources}(\textit{Sources}_C, M, T)$
        \State $\textit{Sinks}_L \gets \textit{LabelSinks}(\textit{Sinks}_C, M, T)$
        \State $A \gets \textit{TaintAnalysis}(P, V, T, \textit{Sources}_L, \textit{Sinks}_L)$
        \State $A \gets \textit{FilterFalsePositives}(A, M, T)$
        \State \textbf{return} $A$
    \end{algorithmic}
\end{algorithm}
}

\vspace{-0.05in}
\subsection{Problem Statement}
\vspace{-0.02in}
We formally define the static taint analysis problem for vulnerability detection. Given a project $P$, taint analysis extracts an inter-procedural data flow graph $\mathbb{G} = (\mathbb{V}, \mathbb{E})$, where $\mathbb{V}$ is the set of nodes representing program expressions and statements, and $\mathbb{E} \subseteq \mathbb{V} \times \mathbb{V}$ is the set of edges representing data or control flow edges between the nodes. A vulnerability detection task comes with two sets $\boldsymbol{V}^C_\textit{source} \subseteq \mathbb{V}$, $\boldsymbol{V}^C_\textit{sink} \subseteq \mathbb{V}$ that denote source nodes where tainted data may originate and sink nodes where a security vulnerability can occur if tainted data reaches it, respectively. Naturally, different classes $C$ of vulnerabilities (or CWEs) have different source and sink specifications. Additionally, there can be sanitizer specifications, $\boldsymbol{V}^C_\textit{sanitizer} \in \mathbb{V}$, that block the flow of tainted data (such as escaping special characters in strings). 

The goal of taint analysis is to find pairs of sources and sinks, ($V_s \in \boldsymbol{V}^C_\textit{source}, V_t \in \boldsymbol{V}^C_\textit{sink}$), such that there is an \emph{unsanitized} path from the source to the sink. More formally,
$\textit{Unsanitized\_Paths}(V_s, V_t) = \exists\text{ }\textit{Path}(V_s, V_t) \text{ s.t. } \forall V_n \in \textit{Path}(V_s, V_t), V_n \notin \boldsymbol{V}^C_\textit{sanitizer}$. Here, $\textit{Path}(V_1, V_k)$ denotes a sequence of nodes $(V_1, V_2, \ldots, V_k)$, such that $V_i \in \mathbb{V}$ and $ \forall i \in 1 \textit{ to } k-1: (v_i, v_{i+1}) \in \mathbb{E}$.

Two key challenges in taint analysis include: 1) identifying relevant taint specifications for each class C that can be mapped to $\boldsymbol{V}^C_\textit{source}$, $\boldsymbol{V}^C_\textit{sink}$ for any project $P$, and 2) effectively eliminating false positive paths in $\textit{Unsanitized\_Paths}(V_s, V_t)$ identified by taint analysis. In the following sections, we discuss how we address each challenge by leveraging LLMs.


\vspace{-0.05in}
\subsection{Candidate Source/Sink Extraction}
\vspace{-0.05in}
\Comment{To perform taint analysis, we require \emph{taint specifications} -- APIs that
return potential sources of malicious user inputs (known as taint
\emph{sources}) and APIs that should not consume such malicious data without proper
sanitization (known as taint \emph{sinks}). Third-party libraries do
not come with such specifications, which makes taint analysis and consequently,
vulnerability detection challenging.}

A project may use various third-party APIs whose specifications may be unknown---reducing the effectiveness of taint analysis. 
In addition, internal APIs might accept untrusted input from downstream libraries.
Hence, our goal is to automatically infer specifications for such APIs. 
We define a specification $S^C$ as a 3-tuple $\langle T, F, R \rangle$, where $T \in \{\textit{ReturnValue}, \textit{Argument}, \textit{Parameter}, \ldots\ \}$ is the type of node to match in $\mathbb{G}$, $F$ is an N-tuple of strings describing the package, class, method name, signature, and argument/parameter position (if applicable) of an API, and $R \in \{\textit{Source}, \textit{Sink}, \textit{Taint-Propagator}, \textit{Sanitizer}\}$ is the role of the API. 
For example, the specification $\langle \textit{Argument}, (\texttt{java.lang}, \texttt{Runtime}, \texttt{exec}, (\texttt{String[]}), 0), \textit{Sink} \rangle$ denotes that the first argument of \texttt{exec} method of \texttt{Runtime} class is a sink for a vulnerability class (OS command injection). A static analysis tool
maps these specifications to sets of nodes $\boldsymbol{V}^C_\textit{source}$ or $\boldsymbol{V}^C_\textit{sink}$ in $\mathbb{G}$.

To identify taint specifications $\boldsymbol{S}_\textit{source}^{C}$ and 
$\boldsymbol{S}_\textit{sink}^{C}$, 
we first extract $\boldsymbol{S}^{\text{ext}}$: external library APIs that are
invoked in the given Java project and are potential candidates to be taint sources or sinks. 
We also extract $\boldsymbol{S}^{\text{int}}$, internal library APIs that are public and may be invoked by a downstream library.
We use CodeQL to extract such candidates and their corresponding metadata such as method name, type signature, enclosing packages and classes, and even JavaDoc documentations, if applicable.

\Comment{Listing~\ref{lst:extractapis} presents our CodeQL query that extracts
external APIs from a Java project.} \Comment{The extracted metadata is useful when
querying LLMs to label the APIs as potential sources or sinks.}

%



\vspace{-0.05in}
\subsection{Inferring Taint Specifications using LLMs}
\vspace{-0.05in}
We develop an automated specification inference technique: $\textit{LabelSpecs}(\boldsymbol{S}^{\#}, \texttt{LLM}, C, R) = \boldsymbol{S}_{R}^{C}$, where $\boldsymbol{S}^{\#} = \boldsymbol{S}^{\text{ext}} \cup \boldsymbol{S}^{\text{int}}$ are candidate specifications for sources and sinks. 
In this work, we do not consider sanitizer specifications, because they typically do not vary for the vulnerability classes that we consider. 
We use LLMs to infer taint specifications.
Specifically, external APIs in $\boldsymbol{S}^{\text{ext}}$ can be classified as either source or sink, while internal APIs in $\boldsymbol{S}^{\text{int}}$ can have their formal parameters identified as sources.
In the Appendix, we show the user prompts for inferring source and sink specifications from external APIs and internal function formal parameters.

Due to the sheer number of APIs to be labeled, we insert a batch of APIs in a single prompt and ask the LLM to respond with JSON formatted strings.
The batch size is a tunable hyper-parameter.
We adopt few-shot (usually 3-shot) prompting strategy for labeling external APIs, while zero-shot is used for labeling internal APIs.
Notably for internal APIs, we also include information from repository readme and JavaDoc documentations, if applicable.
In practice, we find that this extra information helps LLM understand the high-level purpose and usage of the codebase, resulting in better labeling accuracy.
At the end of this stage, we have successfully obtained $\boldsymbol{S}_{\textit{source}}^C$ and $\boldsymbol{S}_{\textit{sink}}^C$ which are going to be used by the static analysis engine in the next stage.

\Comment{Given a project, we consider various candidate specifications that involve external API calls and build specification inference methods for sources and sinks $L_\textit{source}: \boldsymbol{S}_\textit{source}^\textit{candidates} \mapsto \boldsymbol{S}_\textit{source}^{C}$ and $L_\textit{sink}: \boldsymbol{S}_\textit{sink}^\textit{candidates} \mapsto \boldsymbol{S}_\textit{sink}^{C}$, where $C$ is a vulnerability class. For sources, we consider the return values of third-party APIs, and for sinks, we consider the function arguments, as candidates.} 
%

\Comment{
There are a few key challenges in using LLMs for specification inference: First,
LLMs require natural language descriptions of a task (prompts) and have a
limited context window. Given that there may be thousands of external APIs in
the project, how do we design a prompt that can succinctly describe the
specification inference problem, describe the \vulc of interest, and allow LLMs
to effectively label them as sources or sinks? Further, even if some LLMs allow
large contexts (e.g., more than 100,000 tokens), LLMs have been shown to be
particularly unreliable when dealing with information in the middle of long
contexts~\cite{liu2024lost}. Hence, it is impractical to query LLMs using a
single large prompt. \Comment{Second, due to their probabilistic nature LLMs may
sometimes mislabel specifications, reducing their effectiveness. We discuss
how we tackle these challenges through carefully designing prompts and
additional context-based filtering strategies.}
}


\Comment{
In our initial experiments, we found that specifying the signature of
methods produces more consistent results and enables LLMs to accurately infer
the type and purpose of each argument -- which is especially important for sink
inference and handling overloaded methods. Because there may be thousands of
external API calls, we break down the LLM calls into smaller batches. We found
that a batch size of 50 achieves a good balance between context window limits
and result consistency. 

Further, we observed that GPT-4 was very restrictive in its results and only
labeled very few APIs. On closer inspection with an additional explanation
field, we found that GPT-4 often responded that the nature of the API is
context-dependent and hence it cannot be certain of its nature. For instance,
for an API to be a sink, the input data must be malicious. Hence, we append the
user prompt for sinks to assume that the inputs to presented API can be
\emph{potentially malicious}. For sources, we added an additional instruction
specifying that the method may also be invoked on objects that contain
unsanitized user inputs. A common source of this kind is the
\texttt{getRequestURI} method of the \texttt{HttpServletRequest} class that
returns URL components from the protocol name to the query string from a servlet
request object built from a user request. We observed that these instructions
remarkably improved the quality of the inferred specifications.
}

\begin{figure}[t]
    \includegraphics[width=\linewidth]{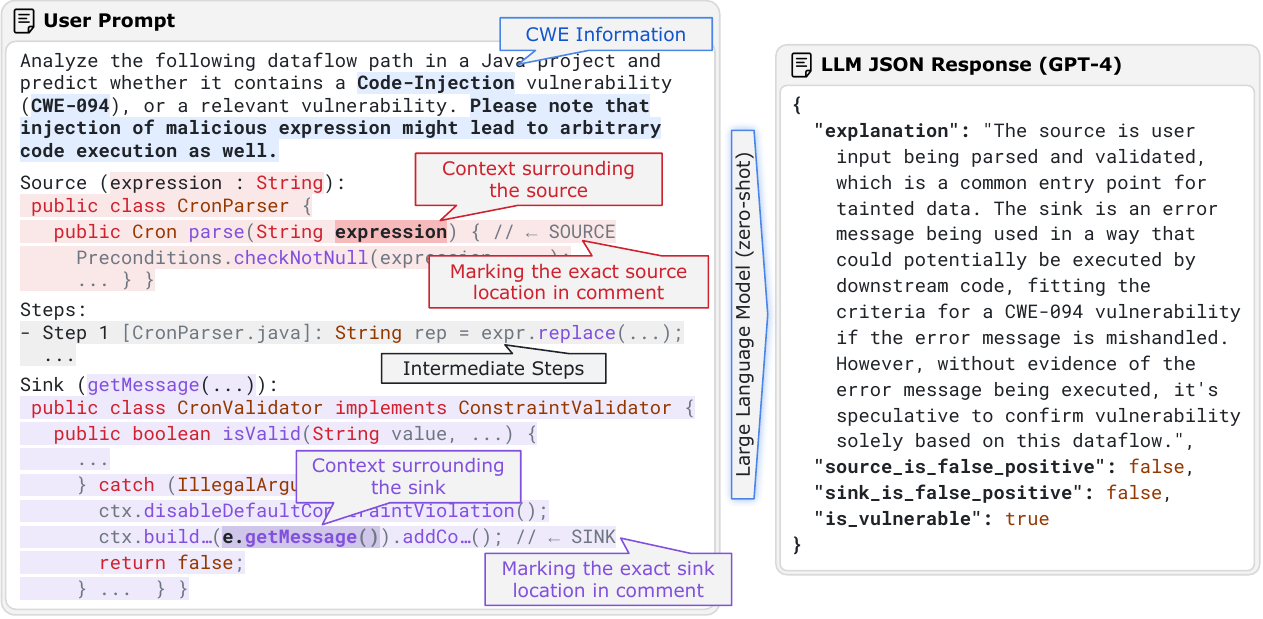}
    \caption{
      LLM user prompt and response for contextual analysis of dataflow paths.
      In the user prompt, we mark with color the CWE and path information that is filling the prompt template.
      For cleaner presentation, we modify the snippets and left out the system prompt.
    }
    \label{fig:contextual-analysis-prompt}
    \vspace{-10px}
\end{figure}

\subsection{Vulnerability Detection}

Once we obtain all the source and sink specifications from the LLM, the next
step is to combine it with a static analysis engine to detect vulnerable paths, i.e., 
$\textit{Unsanitized\_Paths}(V_s, V_t)$, in a given project.
In this work, we use CodeQL~\citep{codeql-web} for this step. CodeQL represents
programs as data flow graphs and provides a query language, akin to Datalog \citep{smaragdakis2010using},
to analyze such graphs. Many security vulnerabilities can be modeled using \emph{queries}
written in CodeQL and can be executed against data flow graphs extracted from
such programs. \Comment{CodeQL supports language-specific database schema and data
structures for languages such as C/C++, Java, Python, and JavaScript.} Given a data flow graph $\mathbb{G}^P$ of a project $P$, CWE-specific
source and sink specifications, and a query for a given vulnerability class $C$, CodeQL returns a set of 
unsanitized paths in the program.
Formally, 
\begin{equation*}
\textit{CodeQL}(\mathbb{G}^{P}, \boldsymbol{S}_\textit{source}^{C}, \boldsymbol{S}_\textit{sink}^C, \textit{Query}^C) = \{\textit{Path}_1, \ldots, \textit{Path}_k\}.
\end{equation*}
\Comment{CodeQL provides specific queries for detecting many common classes of vulnerabilities.}
CodeQL itself contains numerous specifications of third-party
APIs for each \vulc. However, as we show later in our evaluation, despite having
such specialized queries and extensive specifications, CodeQL fails to detect a
majority of vulnerabilities in real-world projects. For our analysis, we write a
specialized CodeQL query for each vulnerability that uses our mined specifications instead of those provided by CodeQL. 
Details of our queries are described in the Appendix.

\subsection{Triaging of Alerts via Contextual Analysis}

Inferring taint specifications only solves part of the challenge. We observe that while LLMs help uncover many new API specifications, sometimes they detect specifications that are not relevant to the \vulc being considered, resulting in too many predicted sources or sinks and consequently many spurious alerts as a result.  For context, even a few hundred taint specifications may sometimes produce thousands of $\textit{Unsanitized\_Paths}(V_s, V_t)$ that a developer needs to triage. 
To reduce the developer burden, we also develop an LLM-based filtering method, $\textit{FilterPath}(\textit{Path}, \mathbb{G}, \texttt{LLM}, C) = \texttt{True}|\texttt{False}$ that classifies a detected vulnerable path (\textit{Path}) in $\mathbb{G}$ as a true or false positive by leveraging context-based and natural language information.



\ifnotNIPSSUB
Fig.~\ref{fig:contextual-analysis-prompt} presents an example prompt for contextual analysis.
The prompt includes CWE information and code snippets for nodes along the path, with an emphasis on the source and sink.
For the intermediate steps, we include the file names and the line of code.
When the path is too long, we keep only a subset of nodes to limit the size of the prompt.
More details and design decisions of this prompt are described in the Appendix.
We expect the LLM to respond in JSON format with the final verdict as well as an explanation to the verdict.
The JSON format prompts the LLM to generate the explanation before delivering the final verdict, as presenting the judgment after the reasoning process is known to yield better results.
In addition, if the verdict is false, we ask the LLM to indicate whether the source or sink is a false positive, which
helps to prune other paths and thereby save on the number of calls to the LLM.

\fi

\Comment{
We expect the LLMs to produce JSON as the response, which includes the final verdict as well as an explanation to the verdict.
The format of the JSON suggests LLM to generate the explanation prior to the final verdict, since giving the judgement after the reasoning process is known to produce better results \todo{Cite Chain-of-thought}.
In addition, if the verdict is false, we ask the LLM to mark whether source or sink is a false positive.
This extra information helps to prune other paths so that we can save on the number of calls to the LLM.}

\Comment{Further, we also develop a User Interface (UX) that
allows developers to quickly triage through the alerts.}


\Comment{
Distinguishing between a true
and false positive result requires understanding the context in which a source
or sink is used. Formally specifying such contextual rules is challenging.
Rather, we leverage LLMs to tackle this task. Our insight here is that providing
sufficient relevant contextual information of a potentially vulnerable code path
and targeting a specific \vulc allows LLMs to better reason about the code
behavior. Listing~\ref{lst:filterprompt} presents our custom LLM prompt template
that we developed for filtering static analysis alarms. We describe how we
populate this template.

CodeQL typically provides a code path starting from a source to a sink with
every security alert. For each such code path, we construct a prompt that
encodes the code path in a summary form. For the source and sink, we extract the
enclosing method and class, and few lines of code before and after the line
containing the source/sink, and include in the prompt. In particular, we replace
\texttt{[SOURCE\_STR]} with the source method and \texttt{[SOURCE\_CODE\_CONTEXT]}
with contextual information of the source. Similarly, we replace the
\texttt{[SINK\_STR]} and \texttt{[SINK\_CODE\_CONTEXT]} placeholder with sink
information. We also include the intermediate nodes in the code path in the
\texttt{[INTERMEDIATE\_STEPS]} block. For each intermediate node at step I, we
include the textual representation of the node in the following format:
\texttt{Step <I> [<FILE>:<FUNCTION>, <NODE\_STRING>]: <NODE\_FULL\_LINE>}, where
\texttt{NODE\_STRING} is the string representation of the node (such as the
method call expression), while the other terms have their usual meaning. Because
some code paths can be very long (up to 50 intermediate steps) and the LLM
context window is limited, we follow a simple heuristic to reduce the prompt
length. For paths longer than 10 intermediate steps, we only include alternate
nodes. In our experiments, we find that this helps limit the prompt length and
provide highly reliable results.
}

\Comment{
\subsubsection{User Interface for Triaging Alerts}
\Fix{we should remove this}
We developed a User Interface that makes it easier for developers to triage
through detected vulnerabilities. While CodeQL provides a plugin to view the
detected results and navigate the detected code paths, we find that it is
cumbersome for a developer to easily sift through many alerts. Further, there is
no way for developer to view similar alerts and eliminate or accept each group
easily. We developed a simple user interface that takes the output of \tool and
populates a Web UI with relevant information about each detected vulnerability.

Figure~\ref{fig:userinterface} shows a snapshot of the user interface for one
project and the details of the detected vulnerabilities. The UI provides
convenient options to mark each alert as a true positive or false positive (via
the \texttt{Is FP?} and \texttt{Is TP?} buttons). By default, the alerts are
grouped based on sink locations. The UI also provides a way to mark similar
alerts, i.e., alerts that share the same source or sinks, as true or false
positives. Developers can also quickly view the source code for each
intermediate node via a pop-up window using the URLs. Finally, the UI provides a
way to download the filtered alerts as a json file for further analysis. We
found that the UI was quite useful for our own analysis and understanding of the
vulnerabilities detected by \tool.

\begin{figure}[!htb]
    \centering
    \frame{\includegraphics[width=1.0\textwidth]{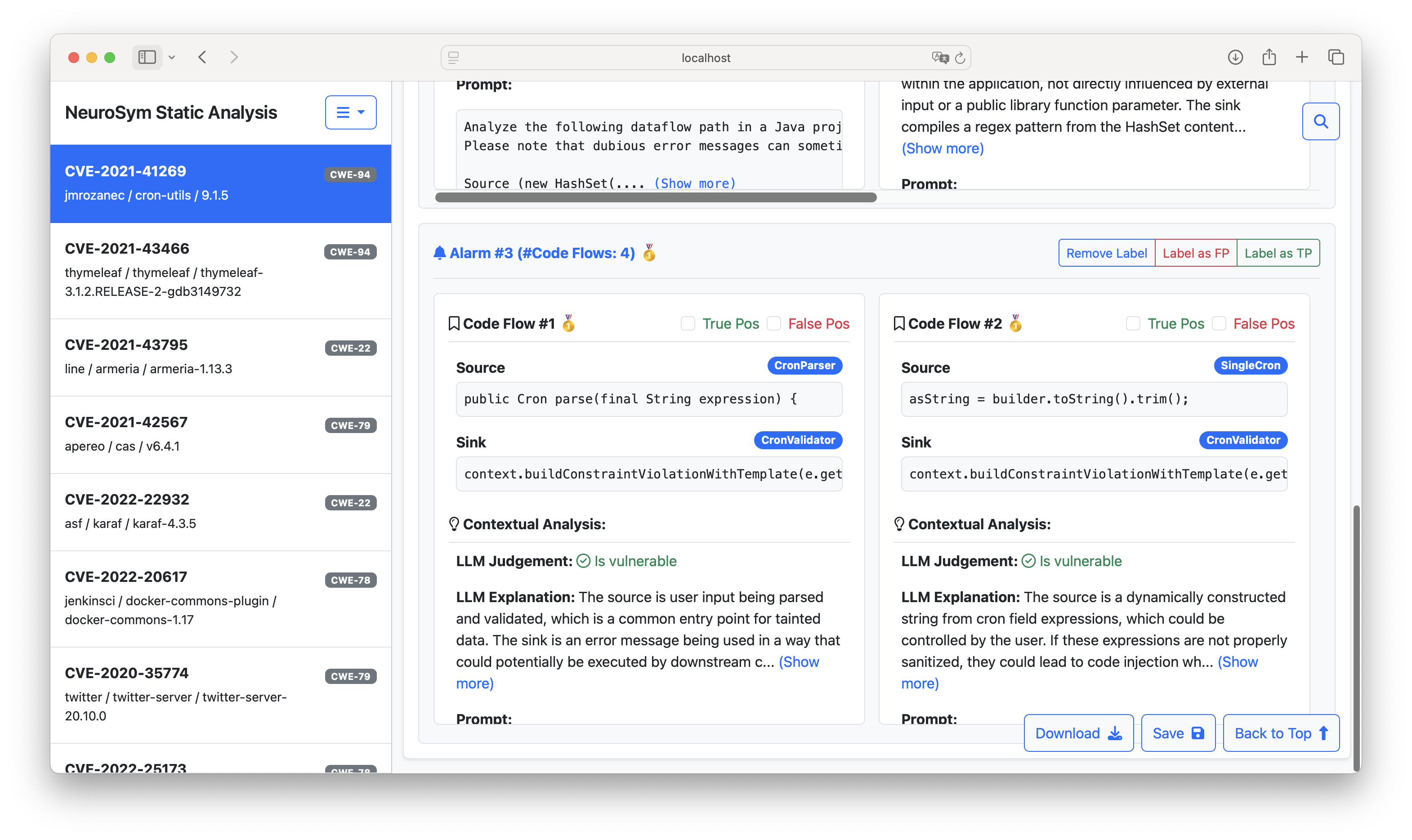}}
    \caption{User Interface for Triaging Alerts}
    \label{fig:userinterface}
\end{figure}
}

\vspace{-0.05in}
\subsection{Evaluation Metrics}
\vspace{-0.02in}

We evaluate the performance of \tool and its baselines using three key metrics: number of vulnerability detected ($\textit{\#Detected}$), average false discovery rate ($\textit{AvgFDR}$), and average F1 ($\textit{AvgF1}$).
For evaluation, we assume that we have a dataset $\mathcal{D} = \{P_1, \dots, P_n\}$ where each $P_i$ is a Java project, and known to contain at least one vulnerability.
The label for a project $P$ is provided as a set of crucial program points $\mathbf{V}_{\text{vul}}^P = \{V_1, \dots, V_n\}$ where the vulnerable paths should pass through, indicated by $\textit{Path} \cap \mathbf{V}_{\text{vul}}^P \neq \emptyset$.
In practice, these are typically the patched methods that can be collected from each vulnerability report.
If at least one detected vulnerable path passes through a fixed location for the given vulnerability, then we consider the vulnerability detected.
Let $\textit{Paths}^P$ be the set of detected paths for each project $P$ from prior stages. The metrics are formally defined as follows:
\begin{align*}
\setlength{\arraycolsep}{1pt}
\begin{array}{rlrl}
\textit{\#VulPath}(P) 
=& 
|\{\textit{Path} \in \textit{Paths}^P ~|~ \textit{Path} ~\cap~ \mathbf{V}_{\text{vul}}^P \neq \emptyset\}|,
&
\textit{Rec}(P)
= &
\mathbbm{1}_{\textit{\#VulPath}(P) > 0},
\\
\textit{\#Detected}(\mathcal{D})
=& 
\textstyle\sum_{P \in \mathcal{D}} \textit{Rec}(P),
&
\textit{Prec}(P)
=&
\textstyle\frac{\textit{\#VulPath}(P)}{|\textit{Paths}^P|},
\\
\textit{AvgFDR}(\mathcal{D}) 
=& 
\text{avg}_{P \in \mathcal{D}, |\textit{Paths}^P|>0} 1 - \textit{Prec}(P),
&
\textit{AvgF1}(\mathcal{D}) 
=& 
\textstyle\frac{1}{|\mathcal{D}|} \textstyle\sum_{P \in \mathcal{D}} \frac{2 \cdot \textit{Prec}(P) \cdot \textit{Rec}(P)}{\textit{Prec}(P) + \textit{Rec}(P)}.
\end{array}
\end{align*}
Specifically, a lower \textit{AvgFDR} is preferable, as it indicates a lower ratio of false positives.
We note that $\textit{Prec}(P)$ might sometimes be undefined due to division-by-zero if the detection tool retrieves no path ($|\textit{Paths}^P| = 0$).
Therefore, for $\textit{AvgFDR}$ to be meaningful, we only consider the projects where at least one positive result is produced ($|\textit{Paths}^P| > 0$).
$\textit{AvgF1}$ avoids this issue since $\textit{Rec}(P) = 0$ when no positive labels exist, forcing the F1 term to be zero regardless of $\textit{Prec}(P)$.



\section{\benchmark: A Dataset of Security Vulnerabilities in Java}

To evaluate our approach, we require a dataset of vulnerable versions of Java projects
with several important characteristics:  1) Each benchmark should have relevant \textbf{vulnerability metadata}, such as the CWE ID, CVE ID, fix commit, and vulnerable project version, 2) each project in the dataset must be \textbf{compilable}, which is a key requirement for static analysis and data flow graph extraction, 3) the projects must be \textbf{real-world}, which are typically more complex and hence challenging to analyze compared to synthetic benchmarks, and 4) finally, each vulnerability and its location (e.g., method) in the project must be \textbf{validated} so that this information can be used for robust evaluation of vulnerability detection tools.
Unfortunately, no existing dataset satisfies all these requirements.

\begin{figure}[t]
\centering
\vspace{-0.1in}
\includegraphics[width=\linewidth]{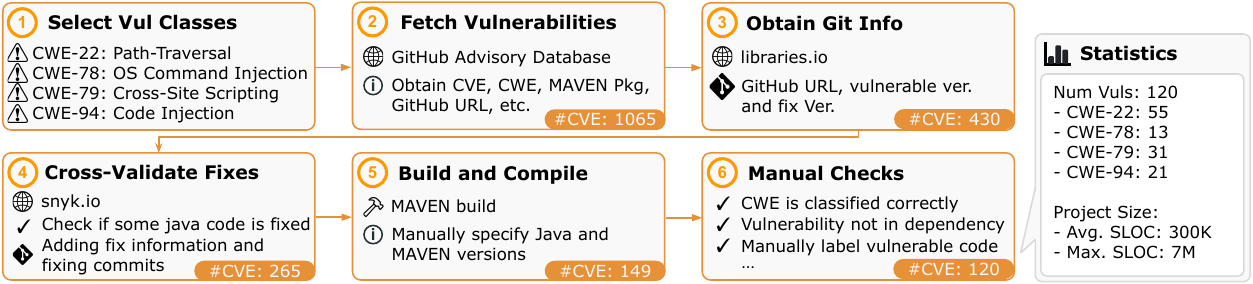}
\caption{Steps for curating \benchmark, and dataset statistics.}
\label{fig:dataset-curation-steps}
\vspace{-0.1in}
\end{figure}


To address these requirements, we curate our own dataset of vulnerabilities. For
this paper, we focus only on vulnerabilities in Java libraries that are
available via the widely used Maven package manager. We choose Java because it
is commonly used to develop server-side, Android, and web applications, which
are prone to security risks. Further, due to Java's long history, there are many
existing CVEs in numerous Java projects that are available for analysis. 
We initially use the GitHub Advisory database~\citep{ghadvdb,ghadvgithub} to obtain such vulnerabilities, and further filter it with cross-validated information from multiple sources, including manual verification. 
\Comment{In the database, each vulnerability is
stored as a Json file with essential metadata such as the CVE ID and CWE ID of
the vulnerability, the vulnerable and fixed versions of the libraries, summary
of the vulnerability, and several links such as the exploit, the fix commits,
and more.} 
Fig.~\ref{fig:dataset-curation-steps} illustrates the complete set of steps for curating \benchmark.

As shown in the statistics (Fig.~\ref{fig:dataset-curation-steps}), the sheer size of these projects make them challenging to analyze for any static analysis tool or ML-based tool. 
Each project in \benchmark comes with GitHub information, vulnerable and fix version, CVE metadata, a script that automatically fetches and builds, and the set of program locations that involve the vulnerability. 

\Comment{
For our work, we focus on four \vulc (or CWEs): \emph{Path Traversal} (CWE
22), \emph{OS Command Injection} (CWE 78), \emph{Cross-Site Scripting or XSS}
(CWE 79), and \emph{Code Injection} (CWE 94). These CWEs are amongst MITRE's top
25 most dangerous CWEs of 2023~\cite{mitretop}. For each CWE, we collect all
CVEs in Java packages from GitHub advisory database. For each CVE, we collect
the CVE ID, the CWE ID, the Maven package name, the vulnerable and fixed package
versions, and the corresponding Github repository URL. The first row in
Table~\ref{tab:datasetstats} presents the initial number of CVEs we obtained in
this step. We observed that the CVEs obtained in this step often did not have
the URL of the corresponding GitHub repository.  Further, the package versions
mentioned in the CVE often did not map directly to a commit or tag in the GitHub
repository. This is important because we require the project source code and the
correct version of code (with the vulnerability) to run our tool. Hence, we use
the \texttt{libraries.io} API~\cite{librariesio} to map the correct GitHub URLs
and version tag names corresponding to each package version obtained from the
CVE data. The second row in Table~\ref{tab:datasetstats} presents the number of
CVEs we successfully mapped in this step. Finally, to determine if we
successfully detect a vulnerability, we require the code location(s) (e.g., the
file or method) that contain the bug, which is typically available in the commit
that fixes the bug. The third row in Table~\ref{tab:datasetstats} presents the
number of CVEs that contain the fix commit information. For this step, we also
manually cross-reference \emph{Snyk}'s database~\cite{synkiodb} to find
additional commits.

Because we use CodeQL for static analysis, we further need to build each project
for CodeQL to extract data flow graphs from the projects. To build each project,
we need to determine the correct Java and Maven compiler versions. We developed
a semi-automated script that tries to build each project with different
combinations of Java and Maven versions. The fourth row in
Table~\ref{tab:datasetstats} presents the number of projects we were able to
build successfully. Overall, this results in \NA{149} projects.

Finally, we manually check each fix commit and validate whether the commit
actually contains a fix to the given CVE in a Java file. For instance, we found
that in some cases the fix is in files written in other languages (such as Scala
or JSP). While code written in other languages may flow to the Java components
in the project during runtime or via compilation, it is not possible to
correctly determine if static analysis can correctly detect such a
vulnerability.  Hence, we exclude such CVEs. Further, we exclude cases where the
vulnerability was in a dependency and the fix was just a version upgrade or if
the vulnerability was mis-classified. Finally, we end up with \NA{120} projects
that we evaluate with \tool. For this task, we divide the CVEs among two
co-authors of the project, who independently validate each case. The co-authors
cross-check each other's results and discuss together to come up with the final
list of projects.}

\vspace{-0.05in}
\section{Evaluation}
\vspace{-0.05in}
\label{sec:evaluation}

We perform extensive experimental evaluations of \tool and demonstrate its practical effectiveness in detecting vulnerabilities in real-world Java repositories in \benchmark.
While additional results and analyses are provided in the Appendix, we address the following key research questions:
\begin{itemize}[leftmargin=*,noitemsep]
    \item \textbf{RQ 1:} How many previously known vulnerabilities can \tool detect?
    \item \textbf{RQ 2:} Does \tool detect new, previously unknown vulnerabilities?
    \item \textbf{RQ 3:} How good are the inferred source/sink specifications by \tool?
    \item \textbf{RQ 4:} How effective are the individual components of \tool?
\end{itemize}

\subsection{Experimental Setup}

We select two closed-source LLMs from OpenAI: GPT-4 (\texttt{gpt-4-0125-preview}) and GPT-3.5 (\texttt{gpt-3.5-turbo-0125}) for our evaluation.
We also select instruction-tuned versions of four open-source LLMs via huggingface API: Llama 3 (L3) 8B and 70B, Qwen-2.5-Coder (Q2.5C) 32B, Gemma-2 (G2) 27B, and DeepSeekCoder (DSC) 7B.
\Comment{To run the open-source LLMs we use two groups of machines: a 2.50GHz Intel Xeon machine, with 40 CPUs, four GeForce RTX 2080 Ti GPUs, and 750GB RAM,
and another 3.00GHz Intel Xeon machine with 48 CPUs, 8 A100s, and 1.5T RAM.}
For the CodeQL baseline, we use version 2.15.3 and its built-in \texttt{Security} queries specifically designed for each CWE.
Other baselines included are Facebook Infer \citep{fbinfer}, SpotBugs \citep{luigi2020spotbugs}, and Snyk \citep{synkio}.
We expand further on the other experimental setups in the Appendix.

\subsection{RQ1: Effectiveness of \tool on Detecting Existing Vulnerabilities}
\label{sec:rq1}

\begin{table}
    \footnotesize
    \centering
    \caption{
        Overall performance comparison of CodeQL vs \tool on Detection Rate ($\uparrow$), Average FDR ($\downarrow$), and Average F1 ($\uparrow$).
        We present results of \tool with LLMs including GPT-4 and GPT-3.5, L3 8B and 70B, Q2.5C 32B, G2 27B, and DSC 7B.
    }
    \label{tab:performance-overall}
    \vspace{-5px}
    \setlength{\tabcolsep}{6pt}
    \begin{tabular}{rl|rrrr}
        \toprule
        & \textbf{Method} & 
        \textbf{\#Detected} (/120) & \textbf{Detection Rate} ($\%$) &
        \textbf{Avg FDR} ($\%$) &
        \textbf{Avg F1 Score}
        \\
        \midrule
        & CodeQL & $27$ & $22.50$ & $90.03$ & $0.076$
        \\
        \midrule
            \multirow{7}{*}{\textbf{IRIS} +} & 
            GPT-4 & 
            $\mathbf{55}$ \green{($\uparrow 28$)} &
            $\mathbf{45.83}$ \green{($\uparrow 23.33$)} & 
            $\mathbf{84.82}$ \green{($\downarrow 5.21$)} & 
            $\mathbf{0.177}$ \green{($\uparrow 0.101$)}
        \\
         & 
            GPT-3.5 & 
            $47$ \green{($\uparrow 20$)} &
            $39.17$ \green{($\uparrow 16.67$)} & 
            $90.42$ \red{($\uparrow 0.39$)} & 
            $0.096$ \green{($\uparrow 0.020$)}
        \\
         & 
            L3 8B & 
            $41$ \green{($\uparrow 14$)} &
            $34.17$ \green{($\uparrow 11.67$)} & 
            $95.55$ \red{($\uparrow 5.52$)} & 
            $0.058$ \red{($\downarrow 0.018$)}
        \\
         & 
            L3 70B & 
            $54$ \green{($\uparrow 27$)} &
            $45.00$ \green{($\uparrow 22.50$)} &
            $90.96$ \red{($\uparrow 0.93$)} &
            $0.113$ \green{($\uparrow 0.037$)}
        \\
         &
            Q2.5C 32B &
            $47$ \green{($\uparrow 20$)} &
            $39.17$ \green{($\uparrow 16.67$)} &
            $92.38$ \red{($\uparrow 2.35$)} &
            $0.097$ \green{($\uparrow 0.021$)}
        \\
         &
            G2 27B &
            $45$ \green{($\uparrow 18$)} &
            $37.50$ \green{($\uparrow 15.00$)} &
            $91.23$ \red{($\uparrow 1.20$)} &
            $0.100$ \green{($\uparrow 0.024$)}
        \\
         & 
            DSC 7B & 
            $52$ \green{($\uparrow 25$)} &
            $43.33$ \green{($\uparrow 20.83$)} & 
            $95.40$ \red{($\uparrow 5.37$)} & 
            $0.062$ \red{($\downarrow 0.014$)}
        \\
        \bottomrule
    \end{tabular}
    \vspace{-5px}
\end{table}

\looseness=-1
The results in Table~\ref{tab:performance-overall} highlight \tool's superior performance compared to CodeQL.
Specifically, \tool, when paired with GPT-4, identifies 55 vulnerabilities—28 more than CodeQL.
While GPT-4 shows the highest efficacy, smaller, specialized LLMs like DeepSeekCoder 7B still detect 52 vulnerabilities, suggesting that our approach can effectively leverage smaller-scale models, enhancing accessibility.
Notably, this increase in detected vulnerabilities does not compromise precision, as evidenced by \tool's lower average false discovery rate (FDR) with GPT-4 compared to CodeQL.
Moreover, \tool improves average F1 by 0.1, reflecting a better balance between precision and recall.
We note that the reported average FDR is a coarse measure as our metrics may consider a true (but unknown) vulnerability found by \tool as a false positive. Hence, the reported FDR is an upper bound.
To get a better sense of detection accuracy, we manually analyzed 50 random alarms reported by \tool (using GPT-4) and found that 27/50 alarms exhibit potential attack surfaces,\emph{ yielding a more refined estimated false discovery rate of 46\%.} Hence, \tool will likely be more effective in practice.

Table~\ref{tab:vul-detect-result-num-detected} presents a detailed breakdown of detected vulnerabilities, comparing \tool against various baselines.
With the exception of \tool using Llama-3 8B, which underperforms in detecting CWE-22 vulnerabilities, \tool consistently outperforms all other baselines.
Notably, CWE-78 (OS Command Injection) remains particularly challenging for all LLMs.
Our manual investigation revealed that the vulnerability patterns in CWE-78 are highly intricate, often involving OS command injections via gadget-chains~\citep{cao2023gadget} or external side effects, such as file writes, which are difficult to track using static analysis.
This highlights the inherent limitations of static analysis, as opposed to dynamic approaches—an area that we leave for future work.

\begin{table}[b]
    \footnotesize
    \centering
    \vspace{-5px}
    \caption{
        Per-CWE statistics of number of vulnerabilities detected ($\textit{\#Detected}$) by baselines and IRIS.
        The compared baselines are CodeQL (QL), Facebook Infer (Infer), Spotbugs (SB), and Snyk. 
        The values in parentheses show the differences of detection by \tool against CodeQL.
    }
    \label{tab:vul-detect-result-num-detected}
    \vspace{-5px}
    \setlength{\tabcolsep}{5pt}
    \begin{tabular}{l|r|rrrr|rrrrr}
        \toprule
            \multirow{2}{*}{\textbf{CWE}} &
            \multirow{2}{*}{\textbf{\#Vuls}} &
            \multicolumn{4}{c|}{\textbf{Baselines}} &
            \multicolumn{5}{c}{\textbf{IRIS} with} 
        \\
        \cmidrule{3-11}
            &
            &
            \textbf{QL} &
            \textbf{Infer} &
            \textbf{SB} &
            \textbf{Snyk} &
            \textbf{GPT-4} &
            \textbf{GPT-3.5} &
            \textbf{L3 8B} &
            \textbf{L3 70B} &
            \textbf{DSC 7B} 
        \\ 
        \midrule
            CWE-22 & 
            55 & 
            22 & 
            0 & 
            2 & 
            21 & 
            \textbf{31} \green{($\uparrow 9$)} & 
            25 \green{($\uparrow 3$)} & 
            19 \red{($\downarrow 3$)} & 
            29 \green{($\uparrow 7$)} & 
            25 \green{($\uparrow 3$)} 
        \\
            CWE-78 & 
            13 & 
            1 & 
            0 & 
            1 & 
            1 & 
            \textbf{3} \green{($\uparrow 2$)} & 
            1 ($= 0$) & 
            2 \green{($\uparrow 1$)} & 
            2 \green{($\uparrow 1$)} & 
            \textbf{3} \green{($\uparrow 2$)} 
        \\
            CWE-79 & 
            31 & 
            4 & 
            0 & 
            1 & 
            1 & 
            13 \green{($\uparrow 9$)} & 
            13 \green{($\uparrow 9$)} & 
            9 \green{($\uparrow 9$)} & 
            \textbf{14} \green{($\uparrow 10$)} & 
            \textbf{14} \green{($\uparrow 10$)} 
        \\
            CWE-94 & 
            21 & 
            0 & 
            0 & 
            0 & 
            0 & 
            8 \green{($\uparrow 8$)} & 
            8 \green{($\uparrow 8$)} & 
            \textbf{11} \green{($\uparrow 11$)} & 
            9 \green{($\uparrow 9$)} & 
            10 \green{($\uparrow 10$)} 
        \\
        \midrule
            \textbf{All} &
            120 & 
            27 & 
            0 & 
            4 & 
            23 & 
            \textbf{55} \green{($\uparrow 28$)} & 
            47 \green{($\uparrow 20$)} & 
            41 \green{($\uparrow 14$)} & 
            54 \green{($\uparrow 27$)} & 
            52 \green{($\uparrow 25$)} 
        \\
        \bottomrule
    \end{tabular}
\end{table}



\Comment{
\NA{The results, as summarized in Table~\ref{tab:vul-detect-result}, showcase the superior performance of \tool{} over CodeQL across both metrics.
Specifically, \tool{} successfully identifies 72 vulnerabilities on a file level and 59 on a method level, outstripping CodeQL by a considerable margin.
This equates to a 33\% (27\%) higher detection rate for \tool{}, underscoring its enhanced ability to uncover existing vulnerabilities within the dataset.}

\NA{To contextualize the significance of these results, we present a detailed examination of a subset of high-profile projects within our dataset in Table~\ref{tab:vul-detect-detail}.
These projects, including notably large codebases such as Nifi with up to 800,000 lines of Java code, represent a challenging benchmark for any vulnerability detection tool.
The analysis extends to the complexity of the software projects, highlighted by an average of 3,500 external APIs for each project under scrutiny.
This detailed look not only illuminates the scale and difficulty of the task at hand but also exemplifies the robustness of \tool{} in navigating complex software environments to detect vulnerabilities effectively.}
}

\subsection{RQ2: Previously Unknown Vulnerabilities by \tool}

We applied \tool with GPT-4 to the latest versions of 30 Java projects.
Among the 16 inspected projects where \tool raised at least one alert, we identified 4 vulnerabilities, including 3 instances of path injection (CWE-22) and one case of code-injection (CWE-94).
To ensure that these vulnerabilities were indeed uncovered due to \tool's integration with LLMs, we verified that CodeQL alone did not detect them.
We highlight one such vulnerability in Fig.~\ref{fig:new-bug}.
CodeQL was unable to detect this issue due to a missing source specification, while GPT-4 successfully flagged the API endpoint \texttt{restoreFromCheckpoint} as a potential entry point for attack.

\subsection{RQ3: Quality of LLM-Inferred Taint Specifications}

\begin{figure}[t]
\begin{minipage}{0.48\linewidth}
\centering
\footnotesize
\includegraphics[width=0.9\linewidth]{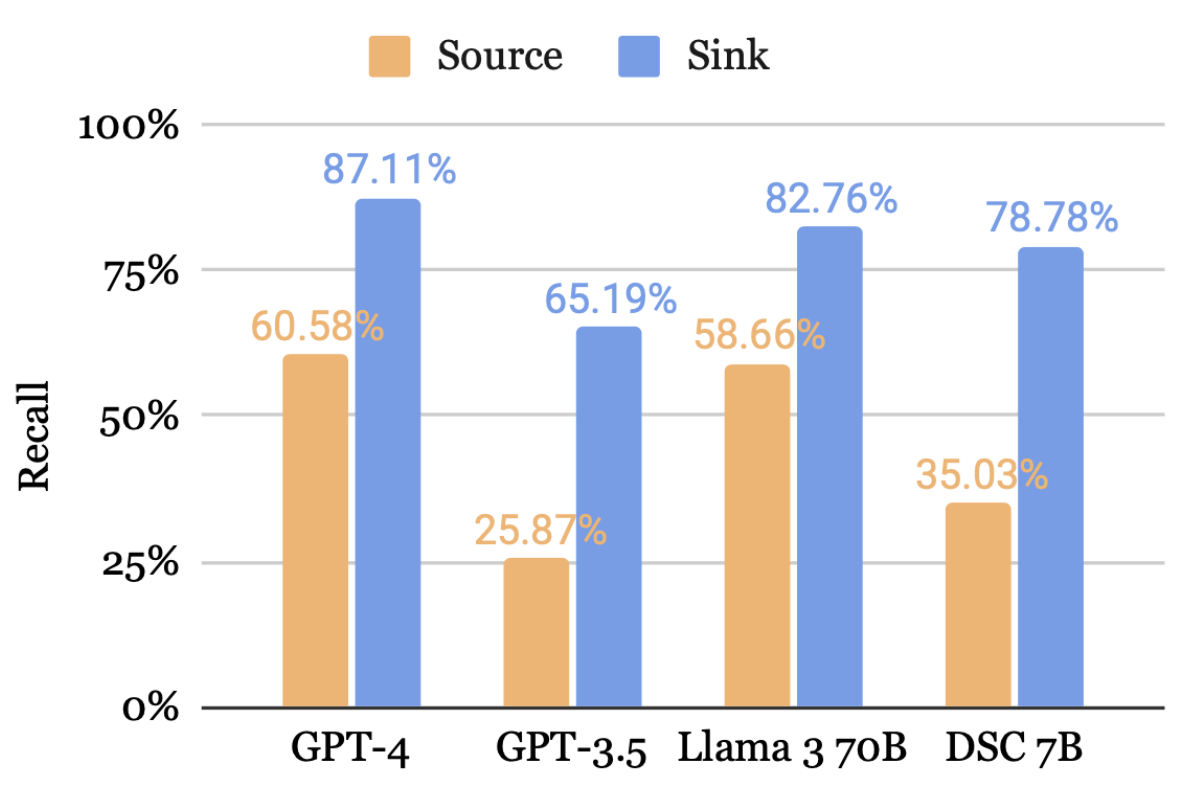}
\vspace{-5px}
\captionof{figure}{Recall of LLM-inferred taint specifications against CodeQL's taint specifications.}
\label{fig:spec-recall-against-codeql}
\end{minipage}
\hfill
\begin{minipage}{0.48\linewidth}
\centering
\footnotesize
\includegraphics[width=0.9\linewidth]{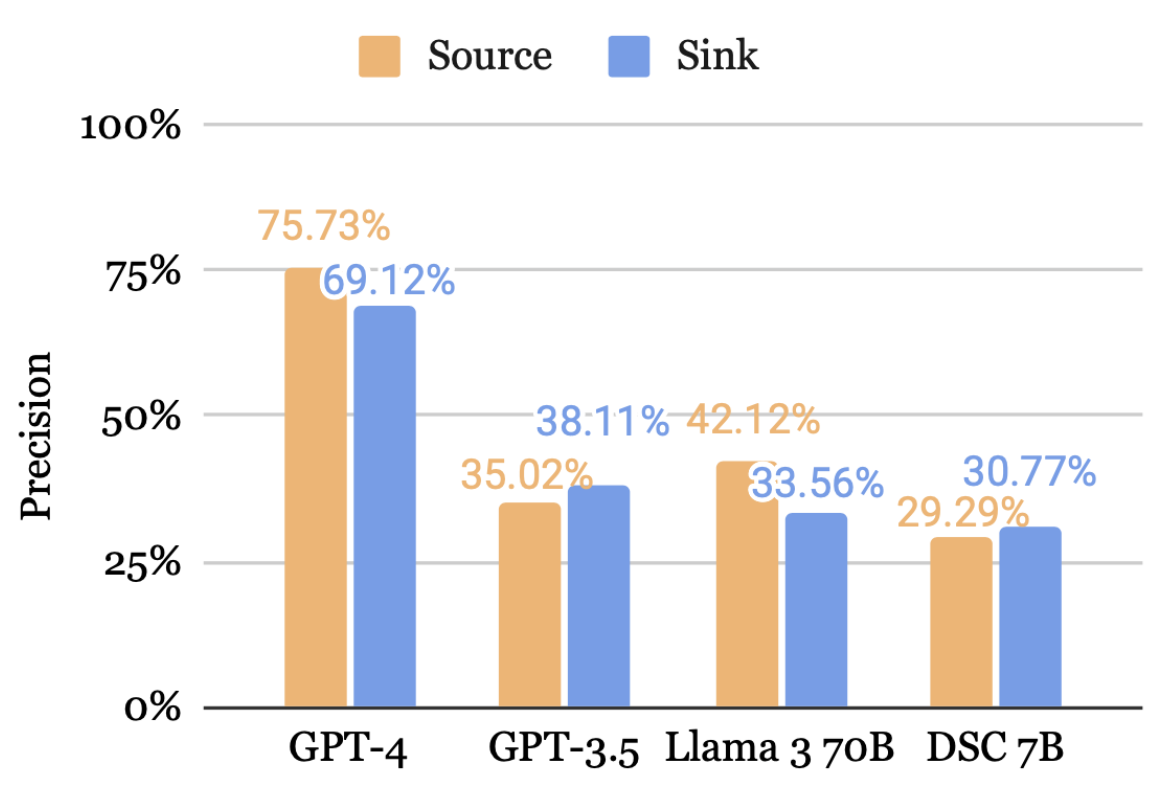}
\vspace{-7px}
\captionof{figure}{Estimated precision of LLM-inferred specifications on randomly sampled labels.}
\label{fig:spec-precision-manual}
\end{minipage}

\vspace{10px}

\begin{minipage}{\linewidth}
\centering
\footnotesize
\includegraphics[width=\linewidth]{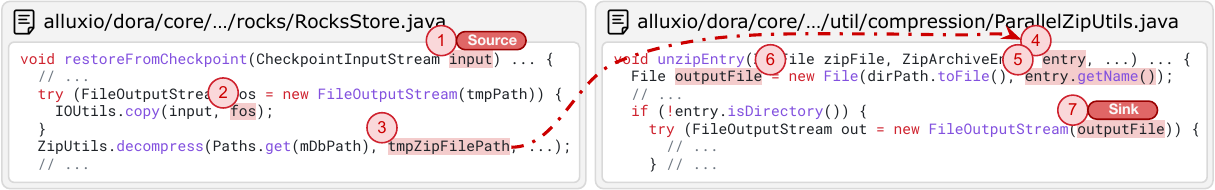}
\captionof{figure}{
A previously unknown vulnerability found in alluxio 2.9.4.
The snippets are slightly modified for presentation purpose.
A user with database restoration permission may supply a database checkpoint Zip file with malicious entry name.
When unzipped, the entry may be written to an arbitrary directory, causing a Zip-Slip vulnerability (CWE-022) that could corrupt the hosting server.
}
\label{fig:new-bug}
\end{minipage}
\vspace{-5px}
\end{figure}

The LLM-inferred taint specifications are fundamental to \tool's effectiveness.
To assess the quality of these specifications, we conducted two experiments.
First, we used CodeQL's taint specifications as a benchmark to estimate the recall of both source and sink specifications inferred by LLMs (Fig.~\ref{fig:spec-recall-against-codeql}).
However, since CodeQL offers a limited set of specifications, we also needed to assess the quality of inferred specifications outside of its known coverage.
To this end, we manually analyzed 960 randomly selected samples of LLM-inferred source and sink labels (30 per combination of CWE and LLM) and estimated the overall precision of the specifications (Fig.~\ref{fig:spec-precision-manual}).

\mypara{LLM-inferred sinks can replace CodeQL sinks}
Overall, LLMs demonstrated high recall when tested against CodeQL's sink specifications (Fig.~\ref{fig:spec-recall-against-codeql}), with GPT-4 scoring the highest ($87.11\%$).
While the recall for source specifications was generally lower, we found that CodeQL tends to over-approximate its source specifications to compensate for a low detection rate.
On the other hand, GPT-4 achieved high precision (over 70\%) in manual evaluations (Fig.~\ref{fig:spec-precision-manual}), aligning with the lower false discovery rate previously reported in Table~\ref{tab:performance-overall}.
For other LLMs, the combination of high recall but lower precision suggests a tendency to over-approximate sink specifications.

\mypara{Over-approximating specifications can benefit IRIS}
Although the precision for LLMs other than GPT-4 is lower, over-approximation can actually help address a core limitation of CodeQL---its restricted set of taint specifications.
By over-approximating, LLMs expand the coverage of taint analysis, offering a partial solution to CodeQL's limited scope.
The impact of this imprecision can be mitigated through contextual analysis as we show next in the ablation studies.

\subsection{RQ4: Ablation Studies}
\label{sec:ablation-studies}

\begin{figure}[t!]

\begin{minipage}{0.46\linewidth}
    \centering
    \footnotesize
    \captionof{table}{
        Ablation on LLM inferred source and sink specifications (CodeQL (QL) versus GPT-4), evaluated using the \textit{\#Detected} metrics.
        When replacing either source or sink with CodeQL specs, we see significantly less vulnerabilities detected.
    }
    \label{tab:ablate-src-sink}
    \vspace{-5px}
    \setlength{\tabcolsep}{3pt}
    \begin{tabular}{l|rrrr|r}
        \toprule
        \textbf{CWE} & \textbf{22} & \textbf{78} & \textbf{79} & \textbf{94} & \textbf{Total}
        \\
        \midrule
      $\text{Src}_{\text{QL}}$ $+$ $\text{Snk}_{\text{QL}}$       &  22 &    1 &   4 &    0  & 27  \red{($\downarrow 28$)}\\
  $\text{Src}_{\text{GPT4}}$ $+$ $\text{Snk}_{\text{QL}}$   &  28 &     3 &   5 &   0 &  36 \red{($\downarrow 19$)}\\
 $\text{Src}_{\text{QL}}$ $+$ $\text{Snk}_{\text{GPT4}}$     &   10  &    1 &    9 &    4 & 24 \red{($\downarrow 31$)}\\ \midrule
 $\text{Src}_{\text{GPT4}}$ $+$ $\text{Snk}_{\text{GPT4}}$   &   31 &    3 &   13 &   8 & 55 \\
        \bottomrule
    \end{tabular}
\end{minipage}
\hfill
\begin{minipage}{0.50\linewidth}
    \centering
    \footnotesize
    \includegraphics[width=\linewidth]{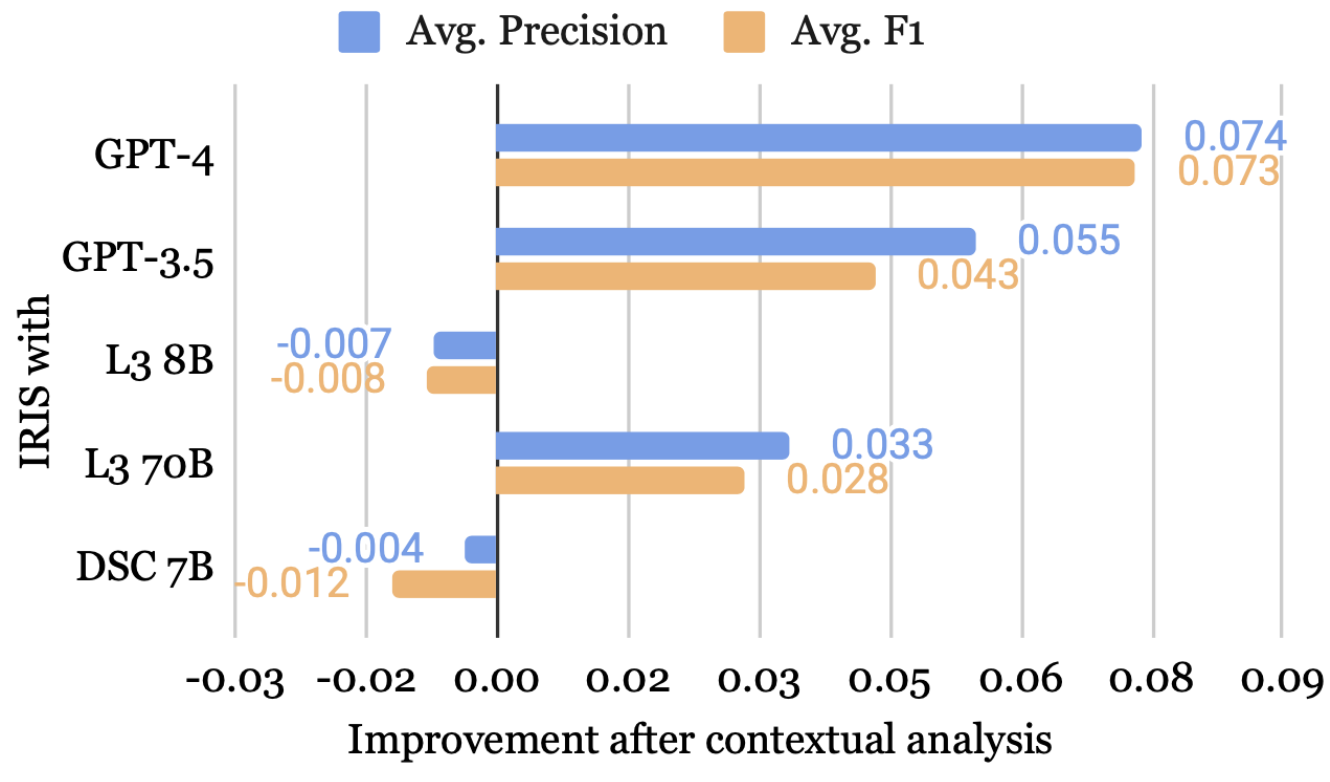}
    \vspace{-15px}
    \captionof{figure}{Improvements of Avg.~Precision and Avg.~F1 after contextual analysis.}
    \label{fig:ablation-contextual-analysis}
\end{minipage}
\vspace{-10px}
\end{figure}

\mypara{Both LLM-inferred sources and sinks are necessary}
Table~\ref{tab:ablate-src-sink} presents additional results when using either only the source or sink specification from an LLM in \tool.
For this experiment, we only use the results with GPT-4 for comparison.
Each row present the number of detected vulnerabilities per CWE.
We observe that omitting either source or sink specifications inferred by GPT-4 causes a drastic reduction in overall recall.

\mypara{Performance gain of contextual analysis depends on LLM's reasoning capability}
As shown in Fig.~\ref{fig:ablation-contextual-analysis}, contextual analysis is highly necessary for the precision and F1 score improvements.
However, only GPT-4, GPT-3.5, and Llama-3 70B see a positive impact after contextual analysis, while the smaller models see negative.
The false positive reduction of contextual analysis is the most effective when the LLM possesses decent reasoning capability.
Indeed, smaller models are more likely to respond with ``vulnerable'' than larger models.

\section{Related Work}
\mypara{Learning-based approaches for vulnerability detection} 
Numerous prior techniques incorporate deep learning for detecting vulnerabilities. This includes Graph Neural Network based models such as \cite{Zhou2019DevignEV, Chakraborty2020DeepLB, Dinella2020Hoppity, Hin2022LineVDSV, Li2021VulnerabilityDW}; LSTM-based models such as \cite{Li2020VulDeeLocatorAD, Li2018SySeVRAF}; and
fine-tuning of Transformer-based models such as \cite{fu2022linevul, steenhoek2023dataflow, Cheng2022PathsensitiveCE}. These approaches focus on method-level detection of vulnerabilities and provide only a binary label classifying a method as vulnerable or not. In contrast, \tool performs whole-project analysis and provides a distinct code path from a source to a sink and can be tailored for detecting different CWEs.
More recently, multiple studies demonstrated that LLMs are not effective at detecting vulnerabilities in real-world code~\citep{steenhoek2024comprehensive,ding2024vulnerability,khare2023understanding}. While these studies only focused on method-level vulnerability detection, it reinforces our motivation that detecting vulnerabilities requires whole-project reasoning, which LLMs currently cannot do alone. 

\mypara{Static analysis tools} 
Apart from CodeQL~\citep{codeql}, other static analysis tools \citep{cppcheck, semgrep, flawfinder, fbinfer, codechecker} also include analyses for vulnerability detection. 
More general query engines \citep{bernhard2016souffle, li2023scallop} have also been applied to find program bugs.
But these tools are not as feature-rich and effective as CodeQL~\citep{li2023comparison,lipp2022empirical}. 
Recently, proprietary tools such as Snyk~\citep{synkio} and SonarQube~\citep{sonarqube} are also gaining in popularity, although sharing the same fundamental limitations of missing specifications and false positives, which \tool improves upon. 
We envision our technique to benefit all such tools.
Works such as MERLIN \citep{livshits2009merlin}, Seldon \citep{chibotaru2019scalable} and InspectJS \citep{dutta2022inspectjs} tackle the problem of specification inference through probabilistic modeling.
Specifically, like \tool,  InspectJS also augments CodeQL with specifications inferred using machine learning. 
However, InspectJS relies on the quality of seed specifications and requires expensive analysis of each third-party library, which \tool does not---making it more scalable. 
Future work could explore incorporating probability estimates for specifications.




\mypara{LLM-based approaches for software engineering} Researchers are increasingly combining LLMs with program reasoning tools for challenging tasks such as fuzzing~\citep{lemieux2023codamosa,xia2024fuzz4all}, program repair~\citep{xia2023automated,joshi2023repair,xia2022less}, and fault localization~\citep{yang2023large}. 
While we are on a similar direction as~\citep{li2024enhancing, wang2024llmdfa}, to our knowledge, our work is among the first to combine LLMs with static analysis to detect application-level security vulnerabilities via whole-project analysis. 
%
\section{Conclusion and Limitations}
We presented \tool, a novel neuro-symbolic approach that combines LLMs with static analysis for vulnerability detection. 
We curate a dataset, \benchmark, containing 120 security vulnerabilities across four classes in real-world projects. 
Our results show that systematically combining LLMs with static analysis significantly improves upon traditional static analysis alone in terms of both detected bugs and the alleviation of developer burden.

\mypara{Limitations} 
There are still many vulnerabilities that \tool cannot detect. 
Future approaches may explore a tighter integration of these two tools to improve performance.
%
In addition, \tool makes numerous calls to LLMs for specification inference and filtering false positives, increasing the potential cost of analysis. 
%
While our results on Java are promising, it is unknown if \tool will perform well on other languages. 
Moreover, there is still a gap between the \tool generated report and the report that the developers would like to see.
We plan to explore this further in future work.
\section*{Acknowledgments}

We thank the anonymous reviewers for their valuable feedback and suggestions that helped improve this work. 
We also thank Claire Wang for helping with open-sourcing the project. 
This research was supported by NSF award CCF 2313010.

\bibliographystyle{iclr2025_conference}
\bibliography{references}

\begin{thebibliography}{52}
\providecommand{\natexlab}[1]{#1}
\providecommand{\url}[1]{\texttt{#1}}
\expandafter\ifx\csname urlstyle\endcsname\relax
  \providecommand{\doi}[1]{doi: #1}\else
  \providecommand{\doi}{doi: \begingroup \urlstyle{rm}\Url}\fi

\bibitem[Avgustinov et~al.(2016)Avgustinov, de~Moor, Jones, and
  Sch{\"a}fer]{codeql}
Pavel Avgustinov, Oege de~Moor, Michael~Peyton Jones, and Max Sch{\"a}fer.
\newblock {QL}: Object-oriented queries on relational data.
\newblock In \emph{European Conference on Object-Oriented Programming}, 2016.

\bibitem[Cao et~al.(2023)Cao, Sun, Wu, Bo, Li, Wu, Liu, He, Ouyang, and
  Li]{cao2023gadget}
Sicong Cao, Xiaobing Sun, Xiaoxue Wu, Lili Bo, Bin Li, Rongxin Wu, Wei Liu,
  Biao He, Yu~Ouyang, and Jiajia Li.
\newblock Improving {Java} deserialization gadget chain mining via
  overriding-guided object generation.
\newblock In \emph{International Conference on Software Engineering}, 2023.

\bibitem[Chakraborty et~al.(2020)Chakraborty, Krishna, Ding, and
  Ray]{Chakraborty2020DeepLB}
Saikat Chakraborty, Rahul Krishna, Yangruibo Ding, and Baishakhi Ray.
\newblock Deep learning based vulnerability detection: Are we there yet?
\newblock \emph{IEEE Transactions on Software Engineering}, 48:\penalty0
  3280--3296, 2020.

\bibitem[Checker Framework()]{checker}
Checker Framework, 2024.
\newblock \url{https://checkerframework.org/}.

\bibitem[Cheng et~al.(2022)Cheng, Zhang, Wang, and
  Sui]{Cheng2022PathsensitiveCE}
Xiao Cheng, Guanqin Zhang, Haoyu Wang, and Yulei Sui.
\newblock Path-sensitive code embedding via contrastive learning for software
  vulnerability detection.
\newblock In \emph{International Symposium on Software Testing and Analysis},
  2022.

\bibitem[Chibotaru et~al.(2019)Chibotaru, Bichsel, Raychev, and
  Vechev]{chibotaru2019scalable}
Victor Chibotaru, Benjamin Bichsel, Veselin Raychev, and Martin Vechev.
\newblock Scalable taint specification inference with big code.
\newblock In \emph{Conference on Programming Language Design and
  Implementation}, 2019.

\bibitem[Code Checker()]{codechecker}
Code Checker, 2023.
\newblock \url{https://github.com/Ericsson/codechecker}.

\bibitem[CPPCheck()]{cppcheck}
CPPCheck, 2023.
\newblock \url{https://cppcheck.sourceforge.io/}.

\bibitem[CVE Trends()]{cvedetails}
CVE Trends, 2024.
\newblock \url{https://www.cvedetails.com}.

\bibitem[Dinella et~al.(2020)Dinella, Dai, Li, Naik, Song, and
  Wang]{Dinella2020Hoppity}
Elizabeth Dinella, Hanjun Dai, Ziyang Li, Mayur Naik, Le~Song, and Ke~Wang.
\newblock Hoppity: Learning graph transformations to detect and fix bugs in
  programs.
\newblock In \emph{International Conference on Learning Representations}, 2020.

\bibitem[Ding et~al.(2024)Ding, Fu, Ibrahim, Sitawarin, Chen, Alomair, Wagner,
  Ray, and Chen]{ding2024vulnerability}
Yangruibo Ding, Yanjun Fu, Omniyyah Ibrahim, Chawin Sitawarin, Xinyun Chen,
  Basel Alomair, David Wagner, Baishakhi Ray, and Yizheng Chen.
\newblock Vulnerability detection with code language models: How far are we?
\newblock \emph{arXiv preprint arXiv:2403.18624}, 2024.

\bibitem[Dutta et~al.(2022)Dutta, Garbervetsky, Lahiri, and
  Sch{\"a}fer]{dutta2022inspectjs}
Saikat Dutta, Diego Garbervetsky, Shuvendu~K Lahiri, and Max Sch{\"a}fer.
\newblock Inspectjs: Leveraging code similarity and user-feedback for effective
  taint specification inference for {Javascript}.
\newblock In \emph{International Conference on Software Engineering: Software
  Engineering in Practice (SEIP) Track}, 2022.

\bibitem[FB Infer()]{fbinfer}
FB Infer, 2023.
\newblock \url{https://fbinfer.com/}.

\bibitem[FlawFinder()]{flawfinder}
FlawFinder, 2023.
\newblock \url{https://dwheeler.com/flawfinder}.

\bibitem[Fu \& Tantithamthavorn(2022)Fu and Tantithamthavorn]{fu2022linevul}
Michael Fu and Chakkrit Tantithamthavorn.
\newblock {LineVul}: A transformer-based line-level vulnerability prediction.
\newblock In \emph{International Conference on Mining Software Repositories},
  2022.

\bibitem[GitHub(2024{\natexlab{a}})]{codeql-web}
GitHub.
\newblock Codeql, 2024{\natexlab{a}}.
\newblock \url{https://codeql.github.com}.

\bibitem[GitHub(2024{\natexlab{b}})]{ghadvdb}
GitHub.
\newblock Github advisory database, 2024{\natexlab{b}}.
\newblock \url{https://github.com/advisories}.

\bibitem[GitHub(2024{\natexlab{c}})]{ghadvgithub}
GitHub.
\newblock Github security advisories, 2024{\natexlab{c}}.
\newblock \url{https://github.com/github/advisory-database}.

\bibitem[Hanam et~al.(2014)Hanam, Tan, Holmes, and Lam]{ranking2014finding}
Quinn Hanam, Lin Tan, Reid Holmes, and Patrick Lam.
\newblock Finding patterns in static analysis alerts: improving actionable
  alert ranking.
\newblock In \emph{International Conference on Mining Software Repositories},
  2014.

\bibitem[He \& Vechev(2023)He and Vechev]{he2023large}
Jingxuan He and Martin Vechev.
\newblock Large language models for code: Security hardening and adversarial
  testing.
\newblock In \emph{Conference on Computer and Communications Security}, 2023.

\bibitem[Heckman \& Williams(2009)Heckman and Williams]{heckman2009model}
Sarah Heckman and Laurie Williams.
\newblock A model building process for identifying actionable static analysis
  alerts.
\newblock In \emph{International Conference on Software Testing, Verification
  and Validation}, 2009.

\bibitem[Hin et~al.(2022)Hin, Kan, Chen, and Babar]{Hin2022LineVDSV}
David Hin, Andrey Kan, Huaming Chen, and Muhammad~Ali Babar.
\newblock Linevd: Statement-level vulnerability detection using graph neural
  networks.
\newblock In \emph{International Conference on Mining Software Repositories},
  2022.

\bibitem[Jimenez et~al.(2023)Jimenez, Yang, Wettig, Yao, Pei, Press, and
  Narasimhan]{jimenez2023swe}
Carlos~E Jimenez, John Yang, Alexander Wettig, Shunyu Yao, Kexin Pei, Ofir
  Press, and Karthik Narasimhan.
\newblock {SWE-Bench}: Can language models resolve real-world github issues?
\newblock \emph{arXiv preprint arXiv:2310.06770}, 2023.

\bibitem[Johnson et~al.(2013)Johnson, Song, Murphy-Hill, and
  Bowdidge]{johnson2013don}
Brittany Johnson, Yoonki Song, Emerson Murphy-Hill, and Robert Bowdidge.
\newblock Why don't software developers use static analysis tools to find bugs?
\newblock In \emph{International Conference on Software Engineering}, 2013.

\bibitem[Joshi et~al.(2023)Joshi, Sanchez, Gulwani, Le, Verbruggen, and
  Radi{\v{c}}ek]{joshi2023repair}
Harshit Joshi, Jos{\'e}~Cambronero Sanchez, Sumit Gulwani, Vu~Le, Gust
  Verbruggen, and Ivan Radi{\v{c}}ek.
\newblock Repair is nearly generation: Multilingual program repair with {LLMs}.
\newblock In \emph{AAAI Conference on Artificial Intelligence}, 2023.

\bibitem[Jung et~al.(2005)Jung, Kim, Shin, and Yi]{jung2005taming}
Yungbum Jung, Jaehwang Kim, Jaeho Shin, and Kwangkeun Yi.
\newblock Taming false alarms from a domain-unaware {C} analyzer by a
  {Bayesian} statistical post analysis.
\newblock In \emph{International Static Analysis Symposium}, 2005.

\bibitem[Kang et~al.(2022)Kang, Aw, and Lo]{kang2022detecting}
Hong~Jin Kang, Khai~Loong Aw, and David Lo.
\newblock Detecting false alarms from automatic static analysis tools: How far
  are we?
\newblock In \emph{International Conference on Software Engineering}, 2022.

\bibitem[Khare et~al.(2023)Khare, Dutta, Li, Solko-Breslin, Alur, and
  Naik]{khare2023understanding}
Avishree Khare, Saikat Dutta, Ziyang Li, Alaia Solko-Breslin, Rajeev Alur, and
  Mayur Naik.
\newblock Understanding the effectiveness of large language models in detecting
  security vulnerabilities.
\newblock \emph{arXiv preprint arXiv:2311.16169}, 2023.

\bibitem[Lavazza et~al.(2020)Lavazza, Tosi, and Morasca]{luigi2020spotbugs}
Luigi Lavazza, Davide Tosi, and Sandro Morasca.
\newblock An empirical study on the persistence of spotbugs issues in
  open-source software evolution.
\newblock In \emph{Quality of Information and Communications Technology}, 2020.

\bibitem[Lemieux et~al.(2023)Lemieux, Inala, Lahiri, and
  Sen]{lemieux2023codamosa}
Caroline Lemieux, Jeevana~Priya Inala, Shuvendu~K Lahiri, and Siddhartha Sen.
\newblock {CODAMOSA}: Escaping coverage plateaus in test generation with
  pre-trained large language models.
\newblock In \emph{International Conference on Software Engineering}, 2023.

\bibitem[Li et~al.(2024)Li, Hao, Zhai, and Qian]{li2024enhancing}
Haonan Li, Yu~Hao, Yizhuo Zhai, and Zhiyun Qian.
\newblock Enhancing static analysis for practical bug detection: An
  llm-integrated approach.
\newblock In \emph{International Conference on Object-Oriented Programming,
  Systems, Languages, and Applications}, 2024.

\bibitem[Li et~al.(2023{\natexlab{a}})Li, Chen, Fan, Feng, Liu, Liu, Liu, and
  Chen]{li2023comparison}
Kaixuan Li, Sen Chen, Lingling Fan, Ruitao Feng, Han Liu, Chengwei Liu, Yang
  Liu, and Yixiang Chen.
\newblock Comparison and evaluation on static application security testing
  ({SAST}) tools for {Java}.
\newblock In \emph{Joint Meeting on European Software Engineering Conference
  and Symposium on the Foundations of Software Engineering},
  2023{\natexlab{a}}.

\bibitem[Li et~al.(2021{\natexlab{a}})Li, Wang, and
  Nguyen]{Li2021VulnerabilityDW}
Yi~Li, Shaohua Wang, and Tien~Nhut Nguyen.
\newblock Vulnerability detection with fine-grained interpretations.
\newblock In \emph{Joint Meeting on European Software Engineering Conference
  and Symposium on the Foundations of Software Engineering},
  2021{\natexlab{a}}.

\bibitem[Li et~al.(2021{\natexlab{b}})Li, Zou, Xu, Jin, Zhu, Chen, Wang, and
  Wang]{Li2018SySeVRAF}
Zhen Li, Deqing Zou, Shouhuai Xu, Hai Jin, Yawei Zhu, Zhaoxuan Chen, Sujuan
  Wang, and Jialai Wang.
\newblock Sysevr: A framework for using deep learning to detect software
  vulnerabilities.
\newblock \emph{IEEE Transactions on Dependable and Secure Computing},
  19:\penalty0 2244--2258, 2021{\natexlab{b}}.

\bibitem[Li et~al.(2020)Li, Zou, Xu, Chen, Zhu, and Jin]{Li2020VulDeeLocatorAD}
Zhuguo Li, Deqing Zou, Shouhuai Xu, Zhaoxuan Chen, Yawei Zhu, and Hai Jin.
\newblock {VulDeeLocator}: A deep learning-based fine-grained vulnerability
  detector.
\newblock \emph{IEEE Transactions on Dependable and Secure Computing},
  19:\penalty0 2821--2837, 2020.

\bibitem[Li et~al.(2023{\natexlab{b}})Li, Huang, and Naik]{li2023scallop}
Ziyang Li, Jiani Huang, and Mayur Naik.
\newblock Scallop: A language for neurosymbolic programming.
\newblock In \emph{Conference on Programming Language Design and
  Implementation}, 2023{\natexlab{b}}.

\bibitem[Lipp et~al.(2022)Lipp, Banescu, and Pretschner]{lipp2022empirical}
Stephan Lipp, Sebastian Banescu, and Alexander Pretschner.
\newblock An empirical study on the effectiveness of static {C} code analyzers
  for vulnerability detection.
\newblock In \emph{International Symposium on Software Testing and Analysis},
  2022.

\bibitem[Livshits et~al.(2009)Livshits, Nori, Rajamani, and
  Banerjee]{livshits2009merlin}
Benjamin Livshits, Aditya~V. Nori, Sriram~K Rajamani, and Anindya Banerjee.
\newblock Merlin: Specification inference for explicit information flow
  problems.
\newblock In \emph{Conference on Programming Language Design and
  Implementation}, 2009.

\bibitem[Pan et~al.(2024)Pan, Ibrahimzada, Krishna, Sankar, Wassi, Merler,
  Sobolev, Pavuluri, Sinha, and Jabbarvand]{pan2024lost}
Rangeet Pan, Ali~Reza Ibrahimzada, Rahul Krishna, Divya Sankar, Lambert~Pouguem
  Wassi, Michele Merler, Boris Sobolev, Raju Pavuluri, Saurabh Sinha, and
  Reyhaneh Jabbarvand.
\newblock Lost in translation: A study of bugs introduced by large language
  models while translating code.
\newblock In \emph{International Conference on Software Engineering}, 2024.

\bibitem[Scholz et~al.(2016)Scholz, Jordan, Suboti\'{c}, and
  Westmann]{bernhard2016souffle}
Bernhard Scholz, Herbert Jordan, Pavle Suboti\'{c}, and Till Westmann.
\newblock On fast large-scale program analysis in {Datalog}.
\newblock In \emph{International Conference on Compiler Construction}, 2016.

\bibitem[Semgrep(2023)]{semgrep}
Semgrep.
\newblock The {Semgrep} platform.
\newblock \url{https://semgrep.dev/}, 2023.

\bibitem[Smaragdakis \& Bravenboer(2010)Smaragdakis and
  Bravenboer]{smaragdakis2010using}
Yannis Smaragdakis and Martin Bravenboer.
\newblock Using {Datalog} for fast and easy program analysis.
\newblock In \emph{International Datalog 2.0 Workshop}, 2010.

\bibitem[Snyk.io()]{synkio}
Snyk.io, 2024.
\newblock \url{https://snyk.io}.

\bibitem[SonarQube()]{sonarqube}
SonarQube, 2024.
\newblock \url{https://www.sonarsource.com/products/sonarqube}.

\bibitem[Steenhoek et~al.(2023)Steenhoek, Gao, and Le]{steenhoek2023dataflow}
Benjamin Steenhoek, Hongyang Gao, and Wei Le.
\newblock Dataflow analysis-inspired deep learning for efficient vulnerability
  detection.
\newblock \emph{arXiv preprint arXiv:2212.08108}, 2023.

\bibitem[Steenhoek et~al.(2024)Steenhoek, Rahman, Roy, Alam, Barr, and
  Le]{steenhoek2024comprehensive}
Benjamin Steenhoek, Md~Mahbubur Rahman, Monoshi~Kumar Roy, Mirza~Sanjida Alam,
  Earl~T Barr, and Wei Le.
\newblock A comprehensive study of the capabilities of large language models
  for vulnerability detection.
\newblock \emph{arXiv preprint arXiv:2403.17218}, 2024.

\bibitem[Wang et~al.(2024)Wang, Zhang, Su, Xu, Xie, and Zhang]{wang2024llmdfa}
Chengpeng Wang, Wuqi Zhang, Zian Su, Xiangzhe Xu, Xiaoheng Xie, and Xiangyu
  Zhang.
\newblock {LLMDFA}: Analyzing dataflow in code with large language models.
\newblock In \emph{Neural Information Processing Systems}, 2024.

\bibitem[Xia \& Zhang(2022)Xia and Zhang]{xia2022less}
Chunqiu~Steven Xia and Lingming Zhang.
\newblock Less training, more repairing please: revisiting automated program
  repair via zero-shot learning.
\newblock In \emph{Joint Meeting on European Software Engineering Conference
  and Symposium on the Foundations of Software Engineering}, 2022.

\bibitem[Xia et~al.(2023)Xia, Wei, and Zhang]{xia2023automated}
Chunqiu~Steven Xia, Yuxiang Wei, and Lingming Zhang.
\newblock Automated program repair in the era of large pre-trained language
  models.
\newblock In \emph{International Conference on Software Engineering}, 2023.

\bibitem[Xia et~al.(2024)Xia, Paltenghi, Le~Tian, Pradel, and
  Zhang]{xia2024fuzz4all}
Chunqiu~Steven Xia, Matteo Paltenghi, Jia Le~Tian, Michael Pradel, and Lingming
  Zhang.
\newblock Fuzz4all: Universal fuzzing with large language models.
\newblock In \emph{International Conference on Software Engineering}, 2024.

\bibitem[Yang et~al.(2023)Yang, Martins, Goues, and Hellendoorn]{yang2023large}
Aidan~ZH Yang, Ruben Martins, Claire~Le Goues, and Vincent~J Hellendoorn.
\newblock Large language models for test-free fault localization.
\newblock \emph{arXiv preprint arXiv:2310.01726}, 2023.

\bibitem[Zhou et~al.(2019)Zhou, Liu, Siow, Du, and Liu]{Zhou2019DevignEV}
Yaqin Zhou, Shangqing Liu, J.~Siow, Xiaoning Du, and Yang Liu.
\newblock Devign: Effective vulnerability identification by learning
  comprehensive program semantics via graph neural networks.
\newblock In \emph{Neural Information Processing Systems}, 2019.

\end{thebibliography}

\appendix


\newpage

\section{Implementation Details of \tool}

\subsection{Selecting Candidate Specifications}
\label{sec:app:candidates}

While extracting external APIs, we filter out commonly-used Java libraries that
are unlikely to contain any potential sources or sinks. Such libraries include
testing libraries like JUnit and Hamcrest or mocking libraries like Mockito.
While we filter out methods that are defined in the project, we specifically allow methods that are inherited from an external class or interface. 
An example is the \texttt{getResource} method of the generic class
\texttt{Class} in \texttt{java.lang} package, which takes a path as a string and accesses a file in the module. Many projects commonly inherit this class and use this method. 
If the input path is unchecked, it may lead to a Path-Traversal vulnerability if the path accesses resources outside the given module. Hence,
detecting such API usages is crucial.  

Taint sources are typically values returned by methods that obtain inputs
from external sources, such as response of an HTTP request or a command line
argument. 
Hence, we select external APIs that have a ``non-void'' return type as candidate sources.
Another type of taint sources are commonly seen in Java libraries.
When used by downstream libraries, tainted information maybe passed into the library through function calls.
Therefore, we also collect the formal parameters for public internal function as source candidates.
Due to the excessive amount of such candidates, we pose a further constraint that the public internal function must be directly invoked by a unit test case within the same repository.
Here, the test cases are identified by checking whether the residing file path has \texttt{src/test} within it.

On the other hand, taint sinks are typically arguments to an external API. 
This involves \textit{explicit} arguments, such as the \texttt{command} argument passed to \texttt{Runtime.exec(String command)} method,
and \textit{implicit} \texttt{this} argument to non-static functions, such as the \texttt{file} variable in the function call \texttt{file.delete()}.
This is the only type of sink that we consider within \tool.

We note that this is not the entire story as there might be other kinds of sources and sinks.
Other types of source candidates include the formal parameter of protected but overridden internal functions (the \texttt{req} parameter in \texttt{protected HTTPServeletResponse doGet(HTTPServeletRequest req)}), arguments to an impure external function (the \texttt{buffer} argument to \texttt{void read(byte[] buffer, int size)}), etc.
Sink candidates include the return value of public facing functions, thrown exceptions, and even static methods without any parameter (\texttt{System.exit()}).
Due to the complexity, we do not tackle such kind of sources of sinks in this work.
However, we plan to explore further in future work.

\subsection{LLM Prompts for Specification Inference}

\begin{figure}[!htb]
\begin{lstlisting}[style=mypromptstyle,caption={LLM prompt for labelling external APIs as sources or sinks.},label={lst:prompt-external}]
System: You are a security expert. You are given a list of APIs to be labeled as potential taint sources, sinks, or APIs that propagate taints. Taint sources are values that an attacker can use for unauthorized and malicious operations when interacting with the system. Taint source APIs usually return strings or custom object types. Setter methods are typically NOT taint sources. Taint sinks are program points that can use tainted data in an unsafe way, which directly exposes vulnerability under attack. Taint propagators carry tainted information from input to the output without sanitization, and typically have non-primitive input and outputs. Return the result as a json list with each object in the format:

{ "package": <package name>,
  "class": <class name>,
  "method": <method name>,
  "signature": <signature of the method>,
  "sink_args": <list of arguments or `this`; empty if the API is not sink>,
  "type": <"source", "sink", or "taint-propagator"> }

DO NOT OUTPUT ANYTHING OTHER THAN JSON.


User: [CWE_LONG_DESCRIPTION]

Some example source/sink/taint-propagator methods are:
[CWE_SOURCE_SINK_EXAMPLES]

Among the following methods, \
assuming that the arguments passed to the given function is malicious, \
what are the functions that are potential source, sink, or taint-propagators to [CWE_TITLE] attack (CWE-[CWE_ID])?

Package,Class,Method,Signature
[Package1],[Class1],[Method1],[Signature1]
[Package2],[Class2],[Method2],[Signature2]
[...]
\end{lstlisting}
\end{figure}

\begin{figure}[!htb]
\begin{lstlisting}[style=mypromptstyle,caption={LLM prompt for labeling formal parameters of internal APIs as sources.},label={lst:prompt-internal}]
System: You are a security expert. You are given a list of APIs implemented in established Java libraries, and you need to identify whether some of these APIs could be potentially invoked by downstream libraries with malicious end-user (not programmer) inputs. For instance, functions that deserialize or parse inputs might be used by downstream libraries and would need to add sanitization for malicious user inputs. On the other hand, functions like HTTP request handlers are typically final and won't be called by a downstream package. Utility functions that are not related to the primary purpose of the package should also be ignored. Return the result as a json list with each object in the format:

{ "package": <package name>,
  "class": <class name>,
  "method": <method name>,
  "signature": <signature>,
  "tainted_input": <a list of argument names that are potentially tainted> }

In the result list, only keep the functions that might be used by downstream libraries and is potentially invoked with malicious end-user inputs. Do not output anything other than JSON.


User: You are analyzing the Java package [PROJECT_AUTHOR]/[PROJECT_NAME]. Here is the package summary:

[PROJECT_README_SUMMARY]

Please look at the following public methods in the library and their documentations (if present). What are the most important functions that look like can be invoked by a downstream Java package that is dependent on [PROJECT_NAME], and that the function can be called with potentially malicious end-user inputs? If the package does not seem to be a library, just return empty list as the result. Utility functions that are not related to the primary purpose of the package should also be ignored.

Package,Class,Method,Doc
[Package1],[Class1],[Method1],[Documentation1]
[Package2],[Class2],[Method2],[Documentation2]
[...]
\end{lstlisting}
\end{figure}

There are two prompts that we use to query LLM for specification inference.
The first one is used to label external APIs as either sources or sinks, illustrated in Listing~\ref{lst:prompt-external}.
At a high level, this is a classification task that classifies each API into one of $\{\textit{Source}, \textit{Sink}, \textit{Taint-Propagator}, \textit{None}\}$.
As shown in the listing, the system prompt involves general instruction about the task and the expected output format, which is JSON.
In the user prompt, we give the description of CWE, since the source and sink specifications of external APIs are dependent on the CWE.
We additionally give few-shot examples that cover both sources and sinks for the given CWE.
At the end, we list out a batch of methods akin to the format of CSV.
Notably for sink specifications, we expect the LLM to give extra information about which exact argument to be considered as the sink.
This include explicit arguments as well as the implicit \texttt{this} argument.
We also note that while taint-propagators are included in the prompt, we do not actually use it in the subsequent stages of \tool.
Primarily, the notion of taint-propagator is to help LLMs differentiate between sinks and summary models, which are sometimes mistakened as sinks.
In general, we find the prompt to serve the purpose well.

The second prompt, depicted in Listing~\ref{lst:prompt-internal}, is used to label the formal parameters of internal APIs as sources.
Since we are analyzing internal API, the information such as project \texttt{README} and function documentations are commonly available.
The goal is to find whether this internal API might be invoked by a downstream library with a malicious input passed to this formal parameter.
This information is not CWE specific, hence no CWE information is included in this prompt.

We hypothesize that since LLMs are pre-trained on internet-scale data, they have knowledge about the behavior of widely used libraries and their APIs.
Hence, it is natural to ask whether LLMs can be used to identify APIs that are
relevant as sources or sinks for any \vulc. 
\Comment{Currently, static analysis tools rely on manually identified specifications, which is labor-intensive and hard to scale, especially given the rapid development and evolution of libraries. } 
If successful, LLMs can alleviate manual effort, and drastically
improve the effectiveness of static analysis tools.

\subsection{Design Decisions of LLM Prompts}

Here are some reasoning behind the design decisions for the LLM prompts:

\begin{enumerate}
\item \textbf{Few-shot for labeling external API}: Since the LLM is tasked with labeling source, sink, and taint-propagator APIs, we provide one example for each category, along with a negative example of an API that does not fall into any of these categories. We typically select examples from the Java standard libraries because they are widely used and their labels are readily available.

\item \textbf{Zero-shot for labeling internal API}: As labels for internal APIs are not available, we rely on the zero-shot capabilities of the language model. To mitigate potential performance loss, we include additional information, such as documentation associated with important internal APIs.

\item \textbf{$\pm 5$ lines surrounding the source and sink location during contextual analysis}: We chose $\pm 5$ lines as a balanced approach to provide sufficient context while managing performance and cost. While technically possible to use a larger window, we observed that excessive context can overwhelm the language model, leading to reduced accuracy. Additionally, a larger context increases computational costs significantly, particularly given the large number of candidate APIs and paths that must be queried.

\item \textbf{Selecting subset of nodes in the alarm path during contextual analysis}: We use a hyperparameter $S$ to control the number of intermediate steps included in the prompt. For paths with more than $S$ intermediate steps, we divide the path into $S$ equal segments and select one step from each. This selection prioritizes function calls, as they may indicate sanitizations. If no function call is present, a node is randomly selected from the segment. In our experiments, we observe that setting $S$ to 10 provides a good balance between the cost and accuracy so that the prompt contains enough context and would not be too long.

\end{enumerate}

\subsection{CodeQL Queries for Static Analysis}

\begin{figure}[!htb]
\begin{lstlisting}[style=codeqlstyle,caption={CodeQL script for detecting vulnerabilities for Path-Traversal (CWE 22).},label={lst:cwe22},escapechar=|]
import java
// other imports ...
import MySources
import MySinks

/**
 * A taint-tracking configuration for tracking flow from remote sources to the 
 * creation of a path.
 */
module MyTaintedPathConfig implements DataFlow::ConfigSig { |\label{lst:line:cwe22:configstart}|
  predicate isSource(DataFlow::Node source) {
    isLLMDetectedSource(source) |\label{lst:line:cwe22:source}|
 }

  predicate isSink(DataFlow::Node sink) {
    isLLMDetectedSink(sink) |\label{lst:line:cwe22:sink}|
  }

  predicate isBarrier(DataFlow::Node sanitizer) {
    sanitizer.getType() instanceof BoxedType or
    sanitizer.getType() instanceof PrimitiveType or
    sanitizer.getType() instanceof NumberType or
    sanitizer instanceof PathInjectionSanitizer
  }

  predicate isAdditionalFlowStep(DataFlow::Node n1, DataFlow::Node n2) {
    isLLMDetectedStep(n1, n2)
  }
} |\label{lst:line:cwe22:configend}|

/** Tracks flow from remote sources to the creation of a path. */
module MyTaintedPathFlow = TaintTracking::Global<MyTaintedPathConfig>;

from MyTaintedPathFlow::PathNode source, MyTaintedPathFlow::PathNode sink
where MyTaintedPathFlow::flowPath(source, sink)
select
  getReportingNode(sink.getNode()),
  source,
  sink,
  "This path depends on a $@.",
  source.getNode(),
  sourceType(source.getNode())
\end{lstlisting}
\end{figure}

Listing~\ref{lst:cwe22} presents our CodeQL query for Path-Traversal vulnerabilities (CWE 22).
Lines~\ref{lst:line:cwe22:configstart}-\ref{lst:line:cwe22:configend} describe a
taint analysis configuration that describes which nodes in the data flow graph
should be considered as sources or sinks. Here, Line~\ref{lst:line:cwe22:source}
specifies our custom predicate \texttt{isLLMDetectedSource} that checks
whether the method called is taint source based on our specifications.
Similarly, our predicates \texttt{isLLMDetectedSink} checks whether the node is a taint sink based on our specifications.
Line~\ref{lst:line:cwe22:sink} checks if a method
call or method argument node is a taint sink based on our specifications. We
generate the source and sink specifications as predicates in QL file as shown in
Listings~\ref{lst:sourcesqll} and~\ref{lst:sinksqll} respectively. Given a
taint configuration and the source and sink specifications, CodeQL can
automatically perform taint analysis on a given project.

We use templates to convert LLM inferred specifications into CodeQL queries.
There are three kinds of queries:
\begin{enumerate}
\item a formal parameter of an internal function as a source;
\item the return value of an external function as a source; and
\item an argument to an external function as a sink.
\end{enumerate}
Example queries for the two kinds of sources are specified in Listing~\ref{lst:sourcesqll}, while the example query for the sink is illustrated in Listing~\ref{lst:sinksqll}.
As shown in the listings, we not only match on function package, class, and name, but also match on individual arguments or parameters.
Moreover, our query handles generic functions or function in generic classes through the \texttt{getSourceDeclaration()} predicate provided by CodeQL.
Notably, when the number of inferred specifications is too large, we will split the single predicate into multiple hierarchical ones, improving the CodeQL performance.

\begin{figure}[!htb]
\begin{lstlisting}[style=codeqlstyle,caption={CodeQL predicate for source specifications.},label={lst:sourcesqll}]
predicate isLLMDetectedSource(DataFlow::Node src) {
    // Sources: Return value from external APIs
    (
        src.asExpr().(Call).getCallee().getName() = "getName" and
        src.asExpr().(Call).getCallee().getDeclaringType().getSourceDeclaration().hasQualifiedName("java.util.zip", "ZipEntry")
    )
    ...
    or
    // Sources: Function formal parameters of internal API
    exists(Parameter p |
        src.asParameter() = p and
        p.getCallable().getName() = "setUserName" and
        p.getCallable().getDeclaringType().getSourceDeclaration().hasQualifiedName("org.apache.dolphinscheduler.dao.entity", "DqRule") and
        ( p.getName() = "userName" )
    )
    ...
}
\end{lstlisting}
\end{figure}

\begin{figure}[!htb]
\begin{lstlisting}[style=codeqlstyle,caption={CodeQL predicate for sink specifications.},label={lst:sinksqll}]
predicate isLLMDetectedSink(DataFlow::Node snk) {
    exists(Call c |
        c.getCallee().getName() = "createTempFile" and
        c.getCallee().getDeclaringType().getSourceDeclaration().hasQualifiedName("java.io", "File") and
        ( c.getArgument(0) = snk.asExpr().(Argument) )
    )
    or
    ...
}
\end{lstlisting}
\end{figure}

\subsection{Visualization of Metrics}

\begin{figure}[t]
  \begin{center}
    \includegraphics[width=\linewidth]{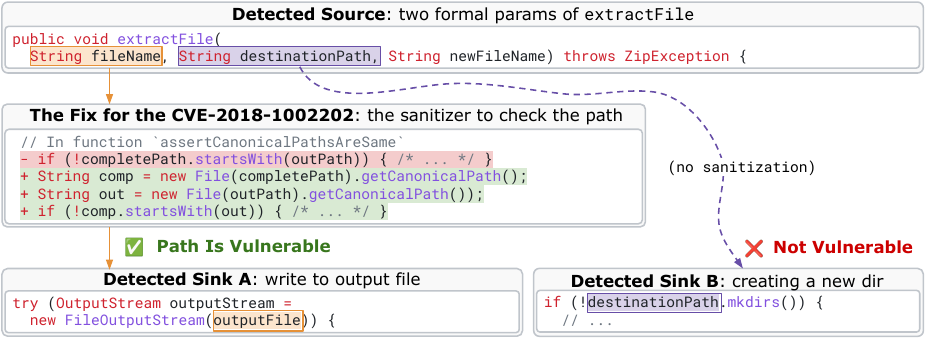}
  \end{center}
  \caption{
    A visualization of our metrics used for vulnerability detection, where the snippets are adapted from Zip4j v2.11.5-2, with slight changes for clearer presentation.
    While both sinks are potential causes of Path-Traversal (CWE-22), only the dataflow path on the left passes through the fixed sanitizer function.
    Therefore, we consider only the path on the left as a vulnerability.
  }
  \label{fig:metrics-visualization}
\end{figure}

We provide a visualization of our $\textit{VulDetected}$ metric in Fig.~\ref{fig:metrics-visualization}.
For evaluation, we assume that the label for a project $P$ is provided as a set of crucial program points $\mathbf{V}_{\text{vul}}^P = \{V_1, \dots, V_n\}$ where the vulnerable paths should pass through.
In practice, these are typically the patched methods that can be collected from each vulnerability report.
As illustrated in Fig.~\ref{fig:metrics-visualization}, if at least one detected vulnerable path passes through a fixed location for the given vulnerability, then we consider the vulnerability detected.
Let $\textit{Paths}^P$ be the set of detected paths for each project $P$ from prior stages. The vulnerable paths inside project $P$ is given by:
\begin{align*}
\setlength{\arraycolsep}{1pt}
\begin{array}{rlrl}
\footnotesize
\textit{VulPaths}(P) 
=& 
\{\textit{Path} \in \textit{Paths}^P ~|~ \textit{Path} ~\cap~ \mathbf{V}_{\text{vul}}^P \neq \emptyset\}
\end{array}
\end{align*}

\section{Additional Details of \benchmark}

\subsection{Details of Dataset Extraction Process}
\label{sec:app:dataset}

\begin{table}[!htb]
    \caption{Vulnerability dataset collection statistics.}
    \centering
    \label{tab:datasetstats}
    \begin{tabular}{l|r|r|r|r|r}\toprule
        \textbf{Step}&\textbf{CWE-22}&\textbf{CWE-78}&\textbf{CWE-79}&\textbf{CWE-94}&\textbf{Total}\\ \midrule
        Initial CVEs&236&39&681&109&1065\\
        W/ Github URL and Version&119&37&219&55&430\\
        W/ Fix Commit&89&27&99&50&265\\
        Compilable&56&17&50&26&149\\
        Fixes in Java Code&56&16&25&47&144\\
        Manual Validation&55&13&31&21&120\\
        \bottomrule
    \end{tabular}
\end{table}

Because we use CodeQL for static analysis, we further need to build each project
for CodeQL to extract data flow graphs from the projects. To build each project,
we need to determine the correct Java and Maven compiler versions. We developed
a semi-automated script that tries to build each project with different
combinations of Java and Maven versions. The fourth row in
Table~\ref{tab:datasetstats} presents the number of projects we were able to
build successfully. Overall, this results in \NA{149} projects.

Finally, we manually check each fix commit and validate whether the commit
actually contains a fix to the given CVE in a Java file. For instance, we found
that in some cases the fix is in files written in other languages (such as Scala
or JSP). While code written in other languages may flow to the Java components
in the project during runtime or via compilation, it is not possible to
correctly determine if static analysis can correctly detect such a
vulnerability.  Hence, we exclude such CVEs. Further, we exclude cases where the
vulnerability was in a dependency and the fix was just a version upgrade or if
the vulnerability was mis-classified. Finally, we end up with \NA{120} projects
that we evaluate with \tool. For this task, we divide the CVEs among two
co-authors of the project, who independently validate each case. The co-authors
cross-check each other's results and discuss together to come up with the final
list of projects.

The closest dataset to ours, in terms of features, is the Java dataset curated
by~\cite{li2023comparison}, containing 165 CVEs. While we initially
considered using their dataset for our work, we found several issues. First,
their dataset does not come with build scripts, which makes it hard to
automatically run each project with CodeQL. Second, their dataset only has few
CVEs for all but one CWE, which makes it difficult to thoroughly analyze a tool
for different vulnerability classes. Finally, they do not provide any automated scripts to
curate more CVEs. Hence, we curated our own dataset and our framework also
allows to easily extend to more vulnerability classes.


\subsection{Comparison of our \benchmark with Existing Vulnerability Datasets}
\label{sec:app:comparison}

We compare \benchmark with existing datasets for vulnerability detection in Java, C, and C++ codebases, on the following criteria:
\begin{enumerate}
\item \textbf{CVE Info}: whether CVE Metadata is contained in the dataset;
\item \textbf{Real-World}: whether the dataset contains real-world projects;
\item \textbf{Fix Locations}: whether the dataset contains fix information about the vulnerabilities;
\item \textbf{Compilable}: whether the dataset ensures that the projects are end-to-end and automatically compilable; and
\item \textbf{Vetted}: whether the vulnerability in the dataset is manually verified and confirmed.
\end{enumerate}
As shown in Table~\ref{tab:datasets}, compared to existing datsets, \benchmark, is the only one that checks every criterion.
This underscores the significance of our new dataset.

\begin{table}[t]
\centering
\setlength{\tabcolsep}{0.2em}
    \caption{Comparison of \benchmark with existing vulnerability datasets.}
    \label{tab:datasets}
    \footnotesize
    \begin{tabular}{l|c|c|c|c|c|c} \toprule
        \textbf{Dataset}&\textbf{Languages}&\textbf{CVE Info}&\textbf{Real-World}&\textbf{Fix Locations}&\textbf{Compilable}&\textbf{Vetted}\\ \midrule
        BigVul&C/C++&\cmark&\cmark&\xmark&\xmark&\xmark\\
        Reveal&C/C++&\xmark&\cmark&\xmark&\xmark&\xmark\\
        CVEFixes&C/C++, Java, ...&\cmark&\cmark&\cmark&\cmark&\xmark\\
        DiverseVul&C/C++&\xmark&\cmark&\cmark&\xmark&\xmark\\
        DeepVD&C/C++&\xmark&\cmark&\xmark&\xmark&\xmark\\
        Juliet&C++, Java&\xmark&\xmark&\cmark&\cmark&\cmark\\
        \cite{li2023comparison}&Java&\cmark&\cmark&\cmark&\xmark&\cmark\\
        SVEN~\citep{he2023large}&C++&\xmark&\cmark&\cmark&\xmark&\cmark\\
         \midrule
        CWE-Bench-Java (Ours) &Java&\cmark&\cmark&\cmark&\cmark&\cmark\\\bottomrule
    \end{tabular}
\end{table}

\section{Evaluation Details}

\subsection{Experimental Settings}

We select two closed-source LLMs from OpenAI: GPT 4 (\texttt{gpt-4-0125-preview}) and GPT 3.5 (\texttt{gpt-3.5-turbo-0125}) for our evaluation.
GPT 4 and GPT 3.5 queries used in the paper are performed through OpenAI API during April and May of 2024.

We also select instruction-tuned versions of six state-of-the-art open-source LLMs via huggingface API: Llama 3 8B and 70B, DeepSeekCoder 7B, QWEN-2.5-Coder 32B, and Gemma-2 27B.
To run the open-source LLMs we use two groups of machines: a 2.50GHz Intel Xeon machine, with 40 CPUs, four GeForce RTX 2080 Ti GPUs, and 750GB RAM, and another 3.00GHz Intel Xeon machine with 48 CPUs, 8 A100s, and 1.5T RAM.

We use CodeQL version 2.15.3 as the backbone of our static analysis.
We have patched CodeQL with an additional feature that augments the Dataflow edge between throw statement and its closest surrounding try-catch block.
We use \href{https://github.com/github/codeql/pull/9914}{this CodeQL pull request} as the base of our patch.


\subsection{CodeQL Baseline}

For baseline comparison with CodeQL, we use the built-in \texttt{Security} queries specifically designed for each CWE that comes with CodeQL 2.15.3.
Note that there are multiple security queries for each CWE, and each produce alarms of different levels (error, warning, and recommendation).
For each CWE, we take the union of alerts generated by all queries and do not differentiate between alarms of different levels.
For instance, there are 3 queries from CodeQL for detecting CWE-22 vulnerabilities, namely \texttt{TaintedPath}, \texttt{TaintedPathLocal}, and \texttt{ZipSlip}.
While \texttt{TaintedPath} and \texttt{ZipSlip} produce error level alarms, \texttt{TaintedPathLocal} produces only alarm recommendations.
To CodeQL's advantage, all alarms are treated equally in our comparisons.

\subsection{Hyper-Parameters and Few-Shot Examples}

During \tool, we have 2 prompts that are used to label external and internal APIs.
Recall that the prompts contain batched APIs.
We use batch size of 20 and 30 for internal and external, respectively.
In terms of few-shot examples passed to labeling external APIs, we use 4 examples for CWE-22, 3 examples for CWE-78, 3 examples for CWE-79, and 3 examples for CWE-94.
We use a temperature of 0, maximum tokens to 2048, and top-p of 1 for inference with all the LLMs. For GPT 3.5 and GPT 4, we also fix a seed to mitigate randomness as much as possible.


\subsection{Details of Selected LLMs}

We include the versions of the 7 selected LLMs in Table~\ref{tab:llmdetails}.

\begin{table}[!htb]
\centering
\footnotesize
\caption{Versions and model IDs of the selected LLMs in our evaluation.}
\label{tab:llmdetails}
\begin{tabular}{ll}
\toprule
\textbf{LLM Version and Size} & \textbf{Model ID}\\ \midrule
GPT 4 & \texttt{gpt-4-0125-preview}\\
GPT 3.5 & \texttt{gpt-3.5-turbo-0125} \\
Llama 3 8B & \texttt{meta-llama/Meta-Llama-3-8B-Instruct} \\
Llama 3 70B & \texttt{meta-llama/Meta-Llama-3-70B-Instruct}\\
DeepSeekCoder 7B & \texttt{deepseek-ai/deepseek-coder-7b-instruct}\\
Qwen-2.5-Coder 32B Instruct & \texttt{Qwen/Qwen2.5-Coder-32B-Instruct}\\
Gemma-2 27B & \texttt{google/gemma-2-27b-it}\\
\bottomrule
\end{tabular}
\end{table}

\subsection{Statistics of Unique and Recurring Specifications}

\begin{table}[!htb]
\centering
\footnotesize
\begin{minipage}{0.48\textwidth}
\centering
\caption{Unique source and sink specifications across all projects in \benchmark.}
\label{tab:unique}
\begin{tabular}{lrrrr}
\toprule 
\textbf{CWE} & \textbf{22} & \textbf{78} & \textbf{79} & \textbf{94} \\ \midrule
\textbf{\#Unique Sources} & 1348 & 899 & 598 & 810 \\
\textbf{\#Unique Sinks} & 1069 & 575 & 514 & 1281 \\
\bottomrule
\end{tabular}
\end{minipage}
\hfill 
\begin{minipage}{0.48\textwidth}
\centering
\centering
\footnotesize
\caption{Recurring source and sink specifications in \benchmark.}
\label{tab:common}
\begin{tabular}{lrrrr}
\toprule 
\textbf{CWE} & \textbf{22} & \textbf{78} & \textbf{79} & \textbf{94} \\ \midrule
\textbf{\#Recurring Sources} & 908 & 232 & 1118 & 626 \\
\textbf{\#Recurring Sinks} & 919 & 201 & 911 & 961 \\
\bottomrule
\end{tabular}
\end{minipage}
\end{table}

\mypara{Continuous taint specification inference is necessary}
Our results show that there is a high number of both unique and recurring sources and sinks.
Table~\ref{tab:unique} presents the number of inferred source and sink specifications that occur only in a single project in \benchmark, whereas Table~\ref{tab:common} presents the specifications that occur in at least two projects.
This indicates that even if previously inferred specifications are useful, a significant number of new relevant APIs still remain and need to be labeled for effective vulnerability detection.
This observation strongly motivates the design of \tool that infers these specifications \emph{on-the-fly} for each project via LLMs, instead of relying on a fixed corpus of specifications like CodeQL.

\subsection{Statistics of Inferred Taint Specifications}

We show the statistics of inferred taint specifications in Table~\ref{tab:source-sink-stats}.
As shown by the percentage, GPT-4 generates smaller set of sources and sinks than smaller-scale LLMs like DeepSeekCoder 7B.

\begin{table}[!htb]
\centering
\footnotesize
\caption{Ratio of API candidates labeled as source (S) or sink (N) by GPT-4 and DeepSeekCoder (DSC) 7B, per CWE and in total.}
\label{tab:source-sink-stats}
\setlength{\tabcolsep}{5pt}
\begin{tabular}{c|r|rr|rr}
        \toprule
        \multirow{2}{*}{\textbf{CWE}}
        & \multirow{2}{*}{\textbf{\#Cand.}}
        & \multicolumn{2}{c|}{\textbf{GPT-4}}
        & \multicolumn{2}{c}{\textbf{DSC 7B}}
        \\ \cmidrule{3-6}
        & & \%S  & \%N & \%S & \%N \\ \midrule
        22 & 130,974 & 2.03\% & 1.90\% & 4.27\% & 4.01\% \\
        78 & 25,605 & 4.73\% & 3.37\% & 3.67\% & 3.33\% \\
        79 & 37,138 & 5.69\% & 4.69\% & 4.28\% & 4.56\% \\
        94 & 36,325 & 5.12\% & 7.83\% & 6.11\% & 6.21\% \\
        \midrule
        \textbf{Total} & 230,042 & 3.41\% & 3.45\% & 4.50\% & 4.37\% \\
        \bottomrule
\end{tabular}
\end{table}

\subsection{Error Analysis: Cause of Undetected Vulnerabilities}

In general, we find the following three main causes of undetected vulnerabilities by employing \tool:

\begin{enumerate}
\item \textbf{Vulnerability cannot be modeled by simple taint dataflows}: for instance, the vulnerability CVE-2020-11977 (CWE-94) is manifested by an unexpected exit(1) call, with no direct taint dataflow going into it. In fact, the taint dataflow goes into the condition of an “if” statement surrounding the exit. In this case, the model of sink API cannot capture the vulnerability.
\item \textbf{Missing dataflow edge due to side-effects}: for instance a taint information is written to a temporary file, and is later read from the same file, subsequently causing a vulnerability. However, the dataflow edge is carried through side-effects, which is not captured by CodeQL.
\item \textbf{Missing dataflow edge due to unspecified library usage}: The vulnerability can only be manifested through a concrete usage of the library; but within the library itself there is no possible dataflow to connect the source and the sink.
\end{enumerate}

Overall, we view the above as general limitations for the static analysis. In terms of LLM induced false negatives, here are the two main failure modes:

\begin{enumerate}
\item \textbf{Missing taint-propagator labels from LLM}: this would cause missing dataflow edges stopping source to flow to sink.
\item \textbf{Missing source or sink specifications from LLM}: if there is no relevant source/sink specification then the static analysis tool would not have the anchor for analysis.
\end{enumerate}

\section{Analysis Runtime}

We include the full table containing statistics to provide more details about projects and our analysis (Table~\ref{tab:longtable}).
For each project, we present its corresponding CWE ID, the lines-of-code (SLOC), the time it takes to run the full analysis, the number candidate APIs and the number of labeled source and sinks by Llama 3 8B.
We also color code cells of interest:
For SLOC, we mark a cell as red if $>$1M; yellow if $>$100k.
For Time, we mark a cell as red if $\geq$1h; yellow if $\geq$5m. 
For the number of candidates, we mark a cell as red if $>$10k.
Lastly for the numbers of sources and sinks, we mark a cell as red if the number is larger than 200.

\setlength{\tabcolsep}{6pt}
\begin{longtable}{llrrrrrr}
\caption{Details of analysis runtime, candidates, and inferred sources and sinks for all projects (Llama 3 8B).}
\label{tab:longtable} \\ \toprule
\textbf{CWE-ID} & \textbf{Project}  & \textbf{SLOC}                 & \textbf{Time}           & \textbf{\#Candidates}              & \textbf{\#Sources}               & \textbf{\#Sinks}                \\ \midrule
\endfirsthead

 22 & DSpace          & \cellcolor{yellow!25}218.2K  & 15s                      & 3.61K                    & 162                   & \cellcolor{red!25}217 \\
    22 & spark           & 10.7K                        & 1m                       & 679                      & 35                    & 27                    \\
    22 & spark           & 9.77K                        & 57s                      & 598                      & 33                    & 22                    \\
    22 & wildfly         & \cellcolor{yellow!25}496.28K & 4m                       & \cellcolor{red!25}14.13K & \cellcolor{red!25}457 & \cellcolor{red!25}425 \\
    22 & vertx-web       & 51.01K                       & 1m                       & 2.06K                    & 80                    & 77                    \\
    22 & camel           & \cellcolor{red!25}1.16M      & \cellcolor{yellow!25}8m  & 293                      & 22                    & 9                     \\
    22 & hutool          & \cellcolor{yellow!25}135.34K & 4m                       & 6.17K                    & 115                   & \cellcolor{red!25}211 \\
    22 & tika            & \cellcolor{yellow!25}106.3K  & 2m                       & 3.84K                    & \cellcolor{red!25}277 & 177                   \\
    22 & retrofit        & 19.28K                       & 1m                       & 880                      & 28                    & 13                    \\
    22 & jspwiki         & \cellcolor{yellow!25}149.45K & 1m                       & 1.83K                    & 62                    & 80                    \\
    22 & camel           & \cellcolor{red!25}1.21M      & \cellcolor{yellow!25}11m & 4.43K                    & 53                    & 80                    \\
    22 & tapestry-5      & \cellcolor{yellow!25}160.06K & 1m                       & 3.04K                    & 91                    & 66                    \\
    22 & spring-cloud-co & 18.56K                       & 1m                       & 1.16K                    & 40                    & 64                    \\
    22 & spring-cloud-co & 18.44K                       & 59s                      & 1.16K                    & 40                    & 64                    \\
    22 & rocketmq        & 94.64K                       & 1m                       & 2.78K                    & 28                    & 54                    \\
    22 & mpxj            & \cellcolor{yellow!25}181.55K & 1m                       & 1.6K                     & 37                    & 43                    \\
    22 & flink           & \cellcolor{red!25}1.14M      & \cellcolor{red!25}2h     & 5.16K                    & 39                    & 61                    \\
    22 & java            & \cellcolor{red!25}1M         & 2m                       & 8.04K                    & 96                    & 41                    \\
    22 & commons-io      & 29.24K                       & 58s                      & 1.07K                    & 12                    & 47                    \\
    22 & karaf           & \cellcolor{yellow!25}135.22K & 1m                       & 5.43K                    & 150                   & \cellcolor{red!25}210 \\
    22 & james-project   & \cellcolor{yellow!25}434.32K & 4m                       & \cellcolor{red!25}14.58K & \cellcolor{red!25}209 & \cellcolor{red!25}226 \\
    22 & vertx-web       & 49.28K                       & 1m                       & 2.36K                    & 83                    & 96                    \\
    22 & esapi-java-lega & 35.26K                       & 59s                      & 1.48K                    & 43                    & 67                    \\
    22 & xwiki-commons   & \cellcolor{yellow!25}103.05K & 1m                       & 3.76K                    & 104                   & 137                   \\
    22 & zip4j           & 16.78K                       & 58s                      & 532                      & 6                     & 34                    \\
    22 & one-java-agent  & 5.19K                        & 51s                      & 327                      & 11                    & 20                    \\
    22 & myfaces         & \cellcolor{yellow!25}161.02K & 1m                       & 2.4K                     & 68                    & 44                    \\
    22 & undertow        & 86.03K                       & 1m                       & 2.58K                    & 66                    & 93                    \\
    22 & DependencyCheck & 28.57K                       & 1m                       & 1.23K                    & 47                    & 66                    \\
    22 & plexus-archiver & 13.04K                       & 51s                      & 573                      & 34                    & 47                    \\
    22 & plexus-archiver & 13.04K                       & 51s                      & 573                      & 34                    & 47                    \\
    22 & zt-zip          & 6.64K                        & 52s                      & 337                      & 14                    & 31                    \\
    22 & curekit         & 511                          & 43s                      & 73                       & 2                     & 4                     \\
    22 & aws-sdk-java    & \cellcolor{red!25}7.72M      & \cellcolor{yellow!25}38m & \cellcolor{red!25}12K    & 62                    & 65                    \\
    22 & venice          & \cellcolor{yellow!25}115.44K & 1m                       & 2.27K                    & 36                    & 79                    \\
    22 & DSpace          & \cellcolor{yellow!25}237.33K & 1m                       & 3.67K                    & 179                   & \cellcolor{red!25}233 \\
    22 & Payara          & \cellcolor{red!25}1.12M      & \cellcolor{yellow!25}7m  & \cellcolor{red!25}16.05K & \cellcolor{red!25}379 & \cellcolor{red!25}427 \\
    22 & DSpace          & \cellcolor{yellow!25}237.33K & 1m                       & 3.67K                    & 179                   & \cellcolor{red!25}233 \\
    22 & goomph          & 12.68K                       & 59s                      & 1.12K                    & 35                    & 111                   \\
    22 & dolphinschedule & 90.69K                       & 1m                       & 3.36K                    & 65                    & 92                    \\
    22 & dolphinschedule & 91.94K                       & 1m                       & 3.4K                     & 65                    & 92                    \\
    22 & testng          & 95.53K                       & 1m                       & 2.08K                    & 33                    & 73                    \\
    22 & uima-uimaj      & \cellcolor{yellow!25}226.81K & 2m                       & 5.66K                    & 103                   & 176                   \\
    22 & keycloak        & \cellcolor{yellow!25}614.82K & \cellcolor{yellow!25}12m & \cellcolor{red!25}13.34K & \cellcolor{red!25}325 & \cellcolor{red!25}252 \\
    22 & glassfish       & \cellcolor{red!25}1.19M      & \cellcolor{yellow!25}5m  & \cellcolor{red!25}12.19K & \cellcolor{red!25}293 & \cellcolor{red!25}346 \\
    22 & graylog2-server & \cellcolor{yellow!25}382K    & 4m                       & \cellcolor{red!25}13.3K  & \cellcolor{red!25}227 & 171                   \\
    22 & mina-sshd       & \cellcolor{yellow!25}130.14K & 1m                       & 3.64K                    & 52                    & 120                   \\
    22 & shiro           & 38.68K                       & 1m                       & 1.5K                     & 41                    & 42                    \\
    22 & plexus-archiver & 15.51K                       & 57s                      & 666                      & 37                    & 56                    \\
    22 & plexus-utils    & 23.3K                        & 58s                      & 754                      & 16                    & 36                    \\
    22 & yamcs           & \cellcolor{yellow!25}693.6K  & 2m                       & \cellcolor{red!25}11K    & 98                    & 113                   \\
    22 & yamcs           & \cellcolor{yellow!25}693.6K  & 2m                       & \cellcolor{red!25}11K    & 98                    & 113                   \\
    22 & shiro           & 38.94K                       & 1m                       & 1.53K                    & 41                    & 43                    \\
    22 & sling-org-apach & 8.34K                        & 54s                      & 695                      & 28                    & 25                    \\
    78 & xstream         & 43.49K                       & 1m                       & 1.39K                    & 91                    & 30                    \\
    78 & xstream         & 59.79K                       & 1m                       & 1.64K                    & 107                   & 42                    \\
    78 & xstream         & 52.25K                       & 1m                       & 1.64K                    & 107                   & 43                    \\
    78 & docker-commons- & 2.79K                        & 54s                      & 362                      & 25                    & 20                    \\
    78 & workflow-cps-pl & 17.02K                       & 1m                       & 1.38K                    & 72                    & 61                    \\
    78 & workflow-cps-gl & 4.31K                        & 55s                      & 523                      & 40                    & 38                    \\
    78 & workflow-multib & 3.45K                        & 53s                      & 500                      & 30                    & 30                    \\
    78 & activemq        & \cellcolor{yellow!25}442.42K & 4m                       & 6.34K                    & \cellcolor{red!25}234 & 192                   \\
    78 & plexus-utils    & 22.76K                       & 1m                       & 714                      & 34                    & 17                    \\
    78 & git-client-plug & 16.41K                       & 1m                       & 1.06K                    & 83                    & 50                    \\
    78 & perfecto-plugin & 667                          & 54s                      & 107                      & 5                     & 10                    \\
    78 & nifi            & \cellcolor{yellow!25}915.95K & \cellcolor{yellow!25}11m & \cellcolor{red!25}22.44K & \cellcolor{red!25}894 & \cellcolor{red!25}614 \\
    78 & script-security & 8.17K                        & 1m                       & 678                      & 40                    & 46                    \\
    79 & antisamy        & 6.38K                        & 57s                      & 381                      & 42                    & 33                    \\
    79 & antisamy        & 6.38K                        & 56s                      & 381                      & 42                    & 33                    \\
    79 & jspwiki         & \cellcolor{yellow!25}149.33K & 1m                       & 1.84K                    & 156                   & 110                   \\
    79 & jspwiki         & \cellcolor{yellow!25}149.33K & 1m                       & 1.84K                    & 156                   & 110                   \\
    79 & jspwiki         & \cellcolor{yellow!25}149.33K & 1m                       & 1.84K                    & 156                   & 110                   \\
    79 & jspwiki         & \cellcolor{yellow!25}157.09K & 1m                       & 1.85K                    & 157                   & 110                   \\
    79 & hibernate-valid & 93.6K                        & 1m                       & 2.06K                    & 79                    & 57                    \\
    79 & cxf             & \cellcolor{yellow!25}798.53K & \cellcolor{red!25}1h     & \cellcolor{red!25}16.54K & \cellcolor{red!25}821 & \cellcolor{red!25}756 \\
    79 & xxl-job         & 9.32K                        & 60s                      & 540                      & 42                    & 41                    \\
    79 & json-sanitizer  & 1.47K                        & 52s                      & 67                       & 4                     & 5                     \\
    79 & hawkbit         & \cellcolor{yellow!25}112.09K & 1m                       & 4.07K                    & 144                   & 151                   \\
    79 & nacos           & \cellcolor{yellow!25}203.78K & 2m                       & 4.08K                    & \cellcolor{red!25}201 & 139                   \\
    79 & antisamy        & 4.93K                        & 1m                       & 362                      & 43                    & 34                    \\
    79 & esapi-java-lega & 35.26K                       & 1m                       & 1.48K                    & 107                   & 85                    \\
    79 & antisamy        & 5.14K                        & 1m                       & 377                      & 44                    & 36                    \\
    79 & jolokia         & 29.97K                       & 1m                       & 1.66K                    & 117                   & 97                    \\
    79 & keycloak        & 60.6K                        & 1m                       & 2.1K                     & 170                   & 136                   \\
    79 & cxf             & \cellcolor{yellow!25}722.83K & \cellcolor{yellow!25}15m & \cellcolor{red!25}15.09K & \cellcolor{red!25}766 & \cellcolor{red!25}710 \\
    79 & sling-org-apach & 1.37K                        & 55s                      & 136                      & 4                     & 13                    \\
    79 & DSpace          & \cellcolor{yellow!25}237.33K & 2m                       & 3.67K                    & \cellcolor{red!25}347 & \cellcolor{red!25}320 \\
    79 & keycloak        & \cellcolor{yellow!25}615.6K  & \cellcolor{red!25}3h     & \cellcolor{red!25}13.37K & \cellcolor{red!25}606 & \cellcolor{red!25}461 \\
    79 & keycloak        & \cellcolor{yellow!25}615.6K  & \cellcolor{red!25}3h     & \cellcolor{red!25}13.37K & \cellcolor{red!25}606 & \cellcolor{red!25}461 \\
    79 & xwiki-commons   & \cellcolor{yellow!25}105.92K & 1m                       & 3.94K                    & \cellcolor{red!25}244 & 151                   \\
    79 & xwiki-commons   & \cellcolor{yellow!25}105.94K & 1m                       & 3.94K                    & \cellcolor{red!25}244 & 151                   \\
    79 & xwiki-rendering & 97.01K                       & 1m                       & 1.22K                    & 73                    & 92                    \\
    79 & xwiki-commons   & \cellcolor{yellow!25}106.87K & 1m                       & 3.99K                    & \cellcolor{red!25}254 & 161                   \\
    79 & jspwiki         & \cellcolor{yellow!25}158.7K  & 1m                       & 2.22K                    & 176                   & 126                   \\
    79 & keycloak        & \cellcolor{yellow!25}617.15K & \cellcolor{red!25}4h     & \cellcolor{red!25}14.04K & \cellcolor{red!25}643 & \cellcolor{red!25}479 \\
    79 & xwiki-commons   & \cellcolor{yellow!25}107.09K & 1m                       & 3.03K                    & \cellcolor{red!25}209 & 143                   \\
    79 & jstachio        & 53.02K                       & 54s                      & 792                      & 40                    & 46                    \\
    79 & xwiki-rendering & 97.63K                       & 1m                       & 1.24K                    & 74                    & 92                    \\
    94 & spring-security & 43.9K                        & 1m                       & 1.83K                    & 120                   & 176                   \\
    94 & xstream         & 52.25K                       & 1m                       & 1.64K                    & 111                   & 145                   \\
    94 & cron-utils      & 13.08K                       & 1m                       & 476                      & 13                    & 26                    \\
    94 & struts          & \cellcolor{yellow!25}160.51K & \cellcolor{yellow!25}12m & 4.39K                    & \cellcolor{red!25}301 & \cellcolor{red!25}357 \\
    94 & activemq        & \cellcolor{yellow!25}547.68K & \cellcolor{red!25}1h     & 7.55K                    & \cellcolor{red!25}370 & \cellcolor{red!25}607 \\
    94 & spring-framewor & \cellcolor{yellow!25}666.11K & \cellcolor{yellow!25}45m & \cellcolor{red!25}17.71K & \cellcolor{red!25}688 & \cellcolor{red!25}846 \\
    94 & spring-cloud-ga & 25.56K                       & 1m                       & 2.01K                    & 130                   & 153                   \\
    94 & dubbo           & \cellcolor{yellow!25}175.63K & 2m                       & 6.73K                    & \cellcolor{red!25}342 & \cellcolor{red!25}383 \\
    94 & incubator-dubbo & 96.35K                       & 1m                       & 3.68K                    & 194                   & \cellcolor{red!25}255 \\
    94 & spring-security & 57.34K                       & 1m                       & 2.43K                    & 192                   & \cellcolor{red!25}234 \\
    94 & kubernetes-clie & \cellcolor{yellow!25}806.35K & 3m                       & 2.33K                    & 93                    & 130                   \\
    94 & commons-text    & 24.87K                       & 1m                       & 962                      & 40                    & 47                    \\
    94 & ff4j            & 46.21K                       & 1m                       & 2.39K                    & 133                   & \cellcolor{red!25}274 \\
    94 & spring-boot-adm & 18.29K                       & 1m                       & 1.83K                    & 92                    & 157                   \\
    94 & sqlite-jdbc     & 17.71K                       & 59s                      & 732                      & 50                    & 74                    \\
    94 & nifi            & \cellcolor{yellow!25}993.76K & \cellcolor{yellow!25}25m & 57                       & 2                     & 11                    \\
    94 & rocketmq        & \cellcolor{yellow!25}108.39K & 2m                       & 3.4K                     & 117                   & 164                   \\
    94 & nifi            & \cellcolor{red!25}1.01M      & \cellcolor{yellow!25}27m & 261                      & 27                    & 24                    \\
    94 & rocketmq        & \cellcolor{yellow!25}197.78K & 2m                       & 6.28K                    & \cellcolor{red!25}205 & \cellcolor{red!25}252 \\
    94 & dolphinschedule & \cellcolor{yellow!25}154.95K & 4m                       & 5.78K                    & \cellcolor{red!25}229 & \cellcolor{red!25}353 \\
    94 & dolphinschedule & \cellcolor{yellow!25}154.95K & 4m                       & 5.78K                    & \cellcolor{red!25}229 & \cellcolor{red!25}353 \\
\bottomrule
\end{longtable}


\end{document}